\documentclass[11pt,journal,oneside,onecolumn,draftclsnofoot]{IEEEtran}

\usepackage{amsmath, amssymb,amscd, epsfig}

\usepackage{graphicx}
\usepackage{mathrsfs}
\usepackage{cite}
\usepackage{algorithm}
\usepackage{algorithmic}
\usepackage{color}
\usepackage{multirow}

\begin{document}
\title{Priority Queueing Models for Cognitive Radio Networks with Traffic Differentiation}
\author{Arash Azarfar,~\IEEEmembership{Student Member,~IEEE}, Jean-Fran\c{c}ois Frigon,~\IEEEmembership{Senior Member,~IEEE}, and Brunilde Sans\`o,~\IEEEmembership{Member,~IEEE}%
\thanks{The authors are from the Department of Electrical Engineering, \'{E}cole Polytechnique de Montr\'{e}al, C.P. 6079, succ. centre-ville, Montr\'{e}al, QC, Canada, H3C 3A7 (e-mail: \{arash.azarfar, j-f.frigon, brunilde.sanso\}@polymtl.ca). \newline \indent This research project was supported by NSERC under Grant STPG365205. \newline This work has been submitted to EURASIP journals for possible
publication. Copyright may be transferred without notice, after which this version may no
longer be accessible.}}

%
%
%

\maketitle

\begin{abstract}
In this paper, we present a new queueing model providing the accurate average system time for packets transmitted over a cognitive radio (CR) link for multiple traffic classes with the preemptive and non-preemptive priority service disciplines. The analysis considers general packet service time, general distributions for the channel availability periods and service interruption periods, and a service-resume transmission. We further introduce and analyze two novel priority service disciplines for opportunistic spectrum access (OSA) networks which take advantage of interruptions to preempt low priority traffic at a low cost. Analytical results, in addition to simulation results to validate their accuracy, are also provided and illustrate the impact of different OSA network parameters on the average system time. We particularly show that, for the same average CR transmission link availability, the packet system time significantly increases in a semi-static network with long operating and interruption periods compared to an OSA network with fast alternating operating and interruption periods. We also present results indicating that, due to the presence of interruptions, priority queueing service disciplines provide a greater differentiated service in OSA networks than in traditional networks. The analytical tools presented in this paper are general and can be used to analyze the traffic metrics of most OSA networks carrying multiple classes of traffic with priority queueing service differentiation. 

\end{abstract}

\section{Introduction}
\label {sec:intro}

Opportunistic spectrum access (OSA) is considered  an important technology to address current and predicted exponential traffic growth in wireless networks~\cite{jondral07,mitola09,xiao13,cisco13}. Such growth is predominantly driven by multimedia traffic, such as video streaming~\cite{cisco13}. Thus, it is expected that OSA networks will carry several traffic classes with different quality of service (QoS) requirements and importance.

The research objective of this paper is to obtain analytical tools to analyze traffic metrics, such as the packet system time, for differentiated services in opportunistic spectrum access networks. Such tools are required to evaluate the packet-level impact of OSA network parameters, novel medium access control (MAC) algorithms, channel sensing order strategies, etc. Moreover, those analytical tools can be used as a decision-making process for multimedia MAC algorithms~\cite{azarfar12b}, for OSA networks employing cognitive radio (CR) nodes, which possess learning and decision-making capabilities.

\subsection{Related Work}

Queueing models is the preferred approach to derive analytical results to analyze traffic metrics \cite{wang-l11}. Priority service disciplines, such as the preemptive and non-preemptive service disciplines, are the most common approaches to implement service differentiation in communication networks.
Furthermore, in an OSA network, the CR users must stop transmitting on an operating channel if the channel's primary user (PU) is detected or if the channel quality is unacceptable due, for example, to deep fading or interference. In the queueing model, the operating channel is the server of the queue. To achieve our objective, we must therefore analyze queueing models with priority service disciplines in the presence of frequent queue server interruptions. 

Queueing models with preemptive priority service discipline and interruptions have been previously studied~\cite{federgruen86,avi-itzhak63,takagi91,gaver62,fiems08}. Some of that work considered that interruption periods are server busy periods generated by higher priority classes of traffic. For this approach, the interruption periods are not generally distributed since they depend on the arrival rate and service rate of higher priority classes. Obtaining the interruption period distribution is therefore not always straightforward. Inversely, given an interruption period distribution, it is not easy to find the appropriate arrival and service processes whose busy period has this distribution. The other articles that studied generally distributed interruption periods only considered a single traffic class and only provided bounds.

Few have attempted to provide queueing models for opportunistic spectrum access networks. In~\cite{wang-l11,rashid07,laourine10}, queueing models for an OSA network with a single class of traffic were derived using a similar approach as in~\cite{federgruen86,avi-itzhak63,takagi91,gaver62,fiems08} whereby the server interruption periods for the cognitive radio users are busy periods generated by the preemptive primary traffic. This approach has several major deficiencies. First, those models are limited to exponential operating period length. Also, as  previously discussed, they can not address arbitrary interruption period lengths before the transmission can resume. This is particularly important for OSA networks since the interruption period length depends on several factors such as the MAC policy, the number of available channels, the number of competing CR users, etc. Even for the simple case where CR users wait until the PU releases the channel, the interruption period is not necessarily distributed as the busy period of a PU Poisson traffic~\cite{chen-D09}. The approach of simply considering the interruption periods as a preemption from PU traffic is therefore not accurate and general enough to analyze OSA networks.

In~\cite{azarfar12e} we  addressed several of those problems in a new queueing model for a single class of CR traffic for general operating and interruption period lengths. However, this model was limited to constant service time. In~\cite{li11}, an optimal threshold for the queue length to decide whether a packet should join the queue or not is derived. However, the model is again not general and can not be used to analyze traffic metrics.

To the best of our knowledge~\cite{kim12} is one of the few papers discussing a queueing system with multiple classes of traffic in cognitive radio networks. The authors analyze a T-preemptive scheme and, similarly to the other work on opportunistic spectrum access networks, the queueing analysis does not consider general interruption lengths and it is specific to the  priority service disciplines
considered.

\subsection{Contributions}

In this paper, we consider real server interruptions distinct from the service times for a high priority class as well as general service time. 
We thus present, to the best of our knowledge, the first queueing model providing the accurate average system time for a Poisson packet arrival process with general service time transmitted over a CR link with general interruption periods and exponentially distributed operating periods for both a single traffic class and for multiple traffic classes with the preemptive and non-preemptive priority service disciplines. We also derive an approximate analysis for general operating period distributions. We further introduce two novel priority service disciplines which are specific to OSA networks with service interruptions. In the first novel OSA service discipline that we name \emph{exceptional non-preemptive}, the service is in general non-preemptive except for low priority arrivals in an empty queue during an interruption period, which can be preempted by high priority packet arrivals during the same interruption period. In the second novel OSA service discipline that we name \emph{preemptive in case of failure}, the service is non-preemptive during the operating periods but high priority packets can preempt low priority packets at the end of an interruption period. We provide an accurate analysis for the first novel OSA priority service discipline while approximate results are provided for the second (exact results are derived for the preemptive in case of failure service discipline for exponential service times).
Since no specific assumptions are made regarding the nature of the operating and interruption periods, the results and derivations presented in this paper can be used to analyze the traffic metrics of most OSA networks with different MAC protocols.
The final contributions of this paper are new insights on OSA networks based on the average system time analysis. Particularly, we show that, for the same average CR transmission link availability (ratio between average channel availability period length and average interruption period), the packet system time significantly increases as the operating and interruption periods average length exceeds the packet service time. We also present results emphasizing the critical importance of minimizing the interruption period lengths to minimize the packet system time in OSA networks. Another conclusion that we present is that priority queueing service disciplines provide a greater differentiated service in OSA networks than in traditional networks.

The reminder of the paper is organized as follows. In Section~\ref{sec:queue-model} we present the cognitive radio system and the queue model. In Section \ref{sec:queue-analysis}, an M/G/1/ queue with interruptions and with a single class of CR traffic is discussed. The results are then used in Section \ref{sec:pr-queueing} to analytically solve four priority queueing disciplines in the presence of interruptions. We also present in Section~\ref{app:alt_approach} an alternative approach to analyze the preemptive and non-preemptive disciplines for exponential operating periods.
Analytical and simulation results are presented in Section \ref{sec:sim-analysis} and finally Section \ref{eq:conclusion} concludes the paper with some remarks on future research directions. 

\section{Cognitive Radio Queue Model}
\label{sec:queue-model}

\begin{figure}%
\includegraphics[width=\columnwidth]{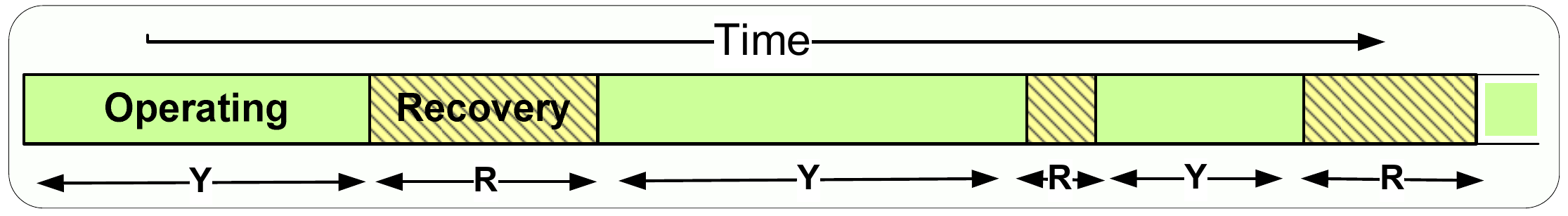}%
\caption{Operation model for a cognitive radio link alternating between operating and recovery (interruption) periods. Identical instances of $Y$ and $R$ are illustrated.}%
\label{fig-queue-channel-model}%
\end{figure}

The cognitive radio queue model can be summarized as follows. We consider a pair of cognitive radio users operating using opportunistic spectrum access over one or more wireless channels. During an operating period, packets that are in one of the CR nodes queue are transmitted to the other CR node according to a chosen service discipline. As illustrated in Fig. \ref{fig-queue-channel-model}, the CR nodes opportunistically operate over a channel for a random duration $Y$ until the channel becomes unavailable. When the channel becomes unavailable, the packet transmission is interrupted for a random length $R$ until an available operating channel can be used by the CR pair, at which time the packet transmission is resumed.

We now describe the details of the model. The OSA network assigns a channel to the pair of CR users according to its MAC protocol and channel assignment algorithm. Transmission over the assigned channel can be multiplexed with other CR users, the only assumption for our model is that during channel availability periods, the pair of CR users have access to a constant service rate over the assigned channel (the packet length defines the distribution of the real service time). If, as in IEEE 802.22, periodic quiet periods are required to sense the channel or perform other OSA network tasks, the channel service rate can be scaled accordingly. Channels are assumed to be homogeneous with the same service rate.
The CR nodes communicate over the assigned channel for a random operating duration $Y$ until the channel becomes unavailable and packet transmissions must be stopped. We denote the instant where the channel becomes unavailable for operation as a failure event~\cite{azarfar11}. To illustrate the generality of this model, we now give a few examples of failure events. A failure can be due to the appearance of the primary user, a false detection of the primary user, a link failure due to excessive transmission loss (fading, shadowing or distance), or interference from other secondary users. A failure event can also be due to the OSA protocol. For example, CR users might have to release a channel after a fixed period of time, even if no primary user appears. 
For the model and its analysis, only the distribution of $Y$ is required and the exact reason for the failure event is irrelevant as long as it is independent of the packet transmission process (e.g., the pair of CR users are not reassigned to a new channel after each packet transmission or when the packet queue is empty).  Note that when the CR users start using a channel, unless it is immediately after a channel unavailability period, they generally have no knowledge about how long this channel has been available. Therefore, $Y$ is a function of the residual time of the availability period of the channel~\cite{azarfar12e}. 

The recovery or interruption period denotes the period of time $R$ during which the CR users can not transmit and try to recover the transmission~\cite{azarfar11}. The length of $R$ depends on the OSA network model but only its distribution is relevant for the queue analysis. We will use a few examples of recovery periods to demonstrate the generality of the proposed model. For OSA MAC protocols in which the CR users buffer the packets until the operating channel becomes available again~\cite{park11}, the distribution of $R$ is identical to the channel unavailability period distribution. For network with a channel switching policy in which when the channel becomes unavailable, the CR users enter a competition with other CR users to be granted access to a new channel~\cite{park11}, the distribution of $R$ will depend on the MAC competition protocol (e.g., slotted Aloha), the number of users, the number of available channels, etc. For OSA networks where a channel is granted by a spectrum server, the length of $R$ can be a fixed period of time (query and service time, radio switching time, etc.).

To summarize, determining the distribution of $Y$ and $R$ according to the OSA network model under study is outside the scope of this paper. But once the distributions are known, the queue model that we are presenting can be used to find the traffic metrics for the OSA network CR users.

\begin{figure}%
\includegraphics[scale=0.65]{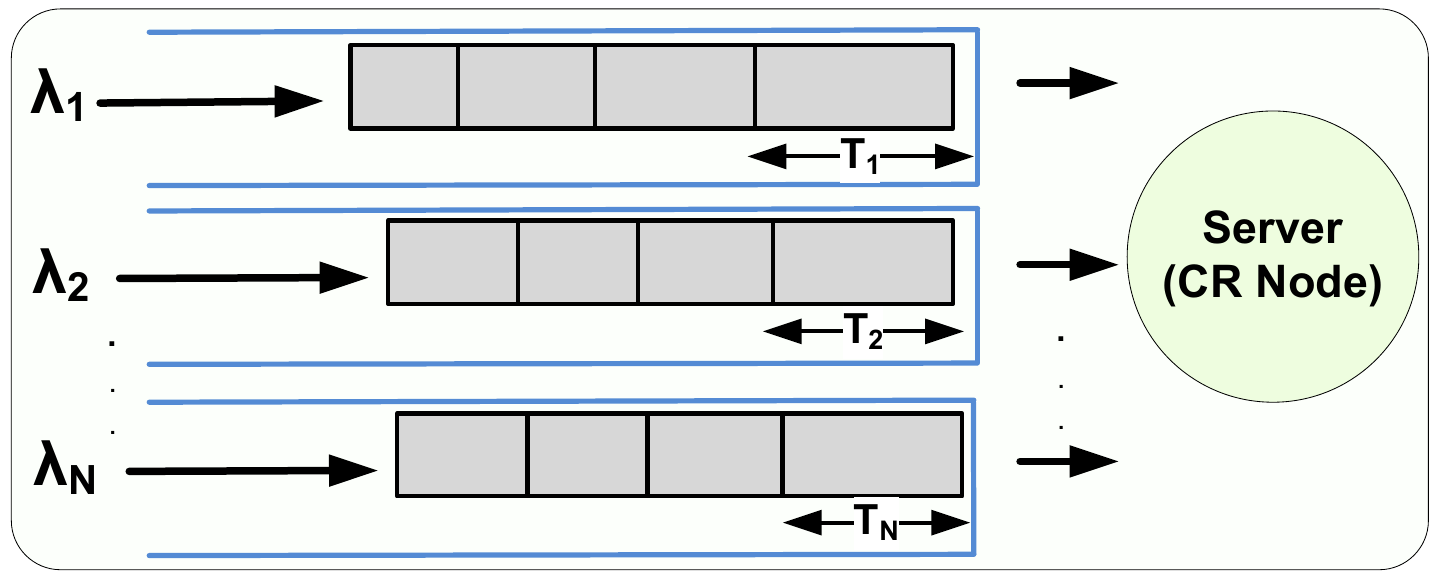}%
\caption{Queue model with a multiple-class cognitive radio traffic.}%
\label{fig-queue-model-Abstract}%
\end{figure}

As illustrated in Fig. \ref{fig-queue-model-Abstract}, we consider a CR system with $N$ traffic classes where each class $i$, $i=1,\dots,N$, has an independent Poisson packet arrival process with rate $\lambda_i$ and the total arrival rate is $\lambda=\sum\lambda_i$. We also denote by $A_i$, the inter-arrival time between packets of class $i$, $i=1,\dots,N$, and define $\mathbb{A} = \min \{ A_1,...,A_{N} \}$ as the inter-arrival time between packets in the system. Throughout the paper, for any random variable $Z$, $f_Z(.)$ and $F_Z(.)$ respectively represent the probability density/mass function (PDF or PMF) and the cumulative distribution function (CDF) of the random variable $Z$. Moreover, $\widehat{Z}(s)$ represents the {Laplaceñ-Stieltjes} transform (LST) of the distribution $F_Z(.)$ of the random variable $Z$. 

Lower index classes have higher priority. In the special case of two traffic classes (e.g., voice and data), we designate the index 1 traffic as high priority (HP) and the index 2 traffic as low priority (LP). Packets from traffic class $i$ have a random \emph{real} service time $T_i$. The real service time is the total transmission time of the packet and excludes the time spent during interruption periods during the service of a packet. From the queueing point of view, the user's operation can thus be modeled as an M/G/1 queue with random service interruptions. 

We must also introduce the notion of \emph{completion time} $X$, which represents the whole time in service for a packet including the real service time $T$ and the interruptions that may occur during its service. We assume a service-resume model which means that after a packet service interruption, only the remaining part of the packet needs to be transmitted. This implies that the completion time of a packet is formed by alternating instances of $Y$ and $R$ named $Y_1, Y_2,\dots$ and $R_1, R_2,\dots$ respectively. 
The queue size is assumed infinite, so the main performance metrics are the total time spent in the queue (waiting time) $W$ and in the system (system time or sojourn time) $D=X+W$. 

\begin{figure}%
\includegraphics[scale=0.65]{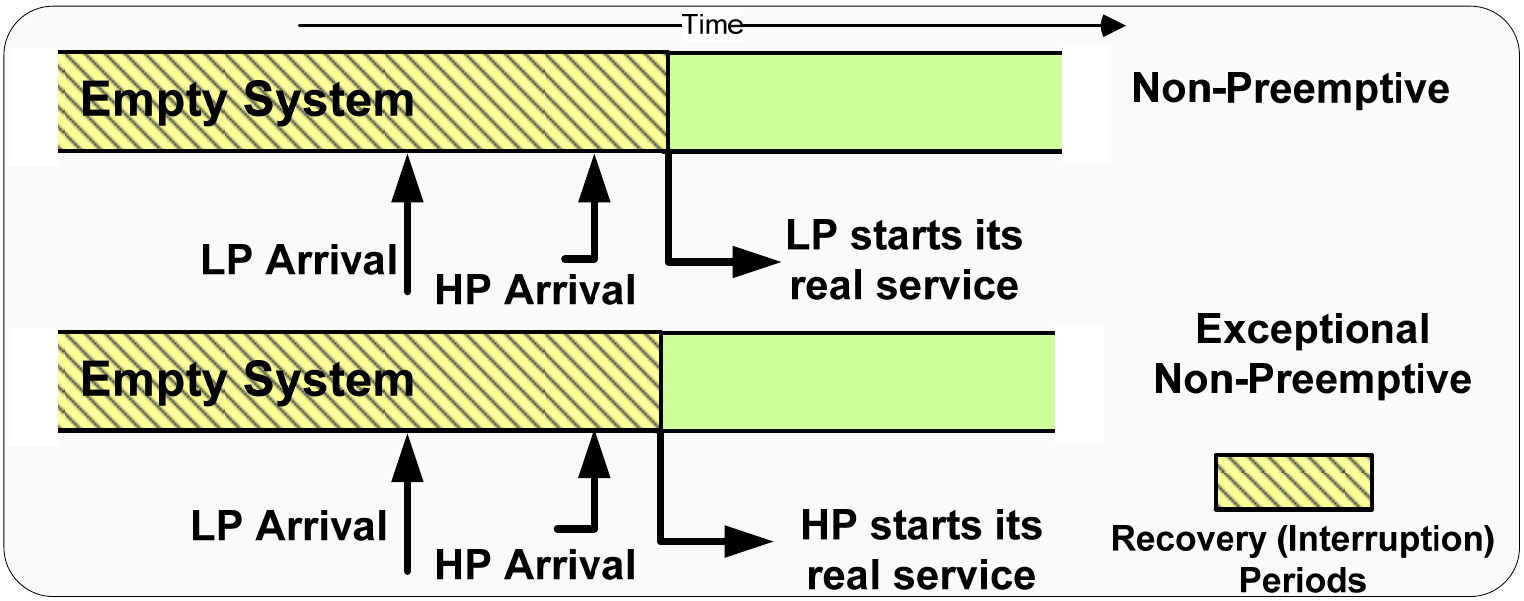}%
\caption{Comparison between non-preemptive and exceptional non-preemptive schemes.}%
\label{Case-U-ENon}%
\end{figure}

We consider four different service disciplines: the classical \emph{non-preemptive} and \emph{preemptive-resume} schemes\cite{bertsekas92}, and two novel disciplines we propose in this paper. As illustrated in Fig.~\ref{Case-U-ENon}, if during a recovery period a low priority (LP) packet arrives in an empty system followed by a high priority (HP) packet, in a non-preemptive scheme the LP packet will be transmitted first. In other words, the LP packet can not be preempted even if its real service has not started yet. 
In the new scheme that we call \emph{exceptional non-preemptive}, an HP packet can preempt a lower priority packet only if its real service has not started yet. As can be seen in Fig.~\ref{Case-U-ENon}, the difference between a non-preemptive and an exceptional non-preemptive scheme is only for the LP packets which arrive to an empty unavailable system. We also propose a \emph{preemption in case of failure} discipline where HP packets can not preempt an LP packet in service until the LP service is finished or if an interruption occurs. In other words, at the end of a recovery period, the priority is always given to HP packets. Meanwhile, in the classical preemptive scheme HP  packets can preempt LP packets at any time. 
The two proposed schemes are defined based on the existence of interruptions and can specifically be used in OSA networks with service interruptions.

\section{Single Traffic Class Analysis}
\label{sec:queue-analysis}

In this section, we analyze the queue model with a single traffic class and obtain results which will be extended in the next section to the analysis of multiple classes of traffic. The results presented here are an extension of the work presented in~\cite{azarfar12e} where only constant service time was considered. In this section, the analysis is done for the general case where the packet length follows an arbitrary random distribution.

\begin{figure}%
\includegraphics[scale=0.65]{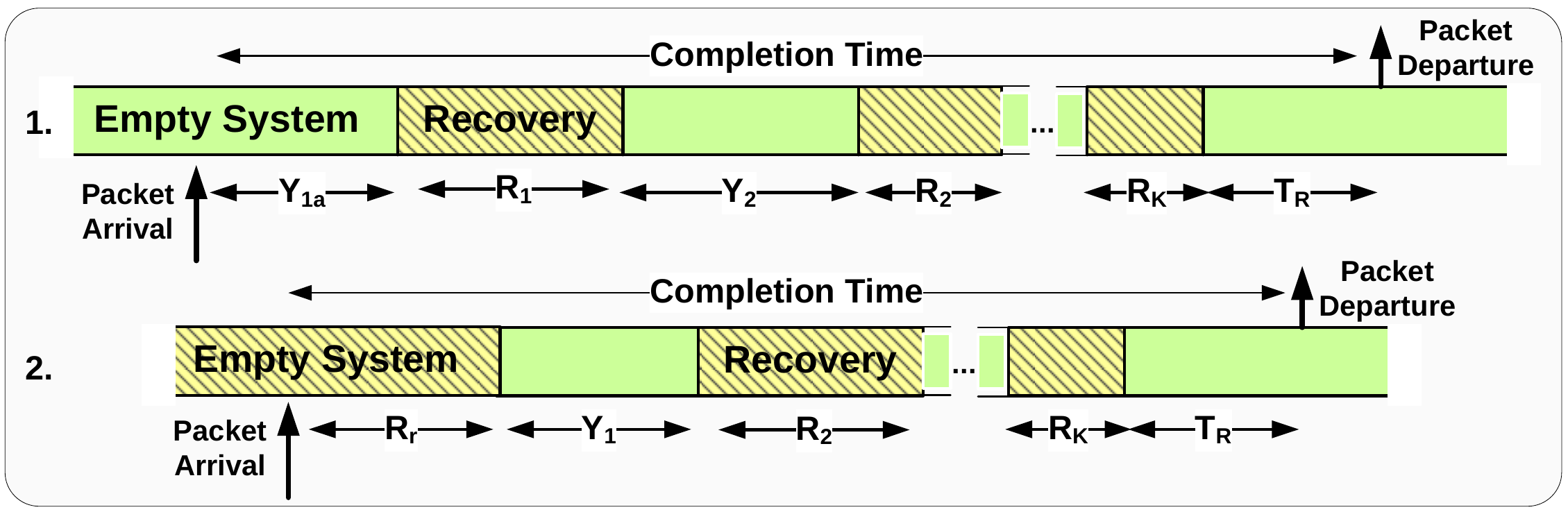}%
\caption{Completion time for the case `$a$' (1.) and  case `$u$' (2.).}%
\label{fig-queue-analyze-Empty}%
\vspace{-0.29cm}
\end{figure}

\begin{figure}%
\includegraphics[scale=0.6]{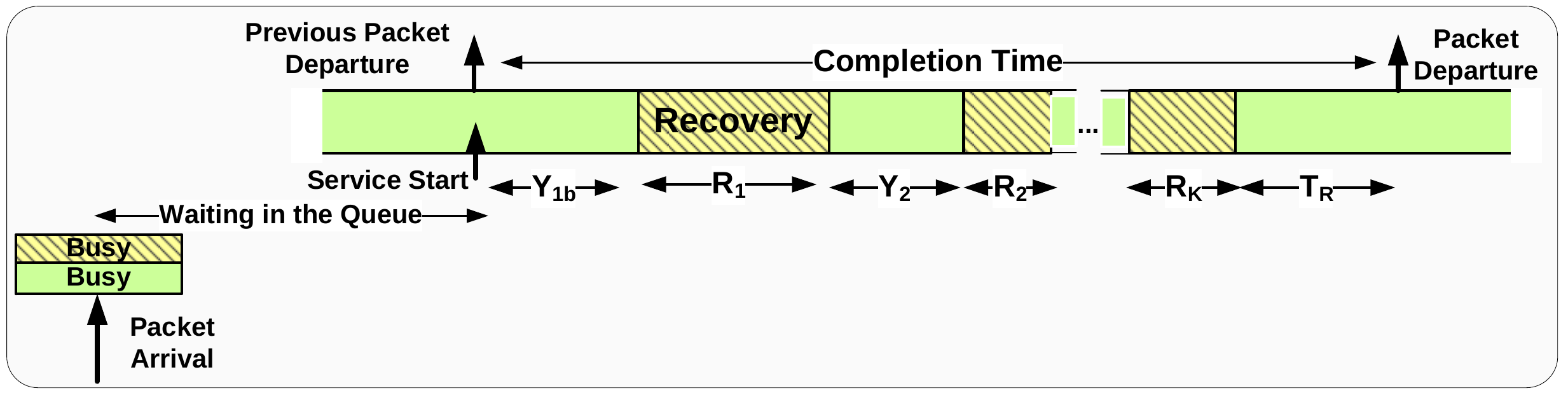}%
\caption{Completion time for the third case when the packet enters a busy system and is queued (case '$b$').}%
\label{fig-queue-analyze-Busy}%
\end{figure}

As can be seen in Fig.'s \ref{fig-queue-analyze-Empty} and \ref{fig-queue-analyze-Busy},  we can distinguish three types of packet in the CR queueing model. 
There are packets which enter an empty available system and their real service starts immediately (see Fig. \ref{fig-queue-analyze-Empty}.1). The subscript `$a$' is used to designate this case. Considering the completion time of the packets of type `$a$', we can see that the distribution of the first operating period is different from the following ones because it represents the residual part of $Y$. We thus use the notation of $Y_{1a}$ to designate the first operating period for the case `$a$' packets. 

Packets that enter an empty system during an interruption period (empty unavailable system) must wait until the end of the recovery period before starting their real service (see Fig. \ref{fig-queue-analyze-Empty}.2). The subscript `$u$' is used to designate this case. For those packets, the distribution of the first operating period is the same as the operating period distribution and is simply denoted by $Y_{1}$. On the other hand, the distribution of the first recovery period is different from the following recovery periods because it represents the residual part of $R$. $R_{r}$ is used to denote the remaining part of the recovery period in which the arrival has occurred. Note that we consider the arrival time as the start of the completion time in  case '$u$' (i.e., $R_{r}$ is not accounted as waiting time but as completion time). 

Finally, there are packets that enter a busy system and are queued (see Fig. \ref{fig-queue-analyze-Busy}). Their service starts immediately after the completion time of the previous packets and the subscript `$b$' is used to designate this case. Note that in this case, the completion time of the packet is always started within an operating period (similar to the case `$a$'). Their first operating period is thus called $Y_{1b}$ because its distribution is different from general $Y$.

In the following, we will find the first two moments of the completion time $X$ and then analyze the average waiting time and other metrics. We also provide simplifications for the special case of exponential operating periods.

\subsection{Completion time}

Suppose $X_a$ represents the completion time of the packets of type `$a$' (similarly $X_b$ and $X_u$ for cases `$b$' and `$u$'). Let also $X_e$ define the completion time of a packet which arrives to an empty system (cases `$a$' and `$u$' together). The first two moments of $X_e$ are given by:
\begin{align}
E[X_e] &=P_{ae}E[X_a] + (1-P_{ae})E[X_u], \\
E[X_e^2] &=P_{ae}E[X_a^2] + (1-P_{ae})E[X_u^2],
\label{eq:}
\end{align}
where $P_{ae}$ is the average probability that the server is available when the system is empty. From \cite{federgruen86}, $P_{ae}$ can be found equal to:
\begin{equation}
P_{ae}=1-\frac{(1-\widetilde{F}_Y(\lambda))(1-\widetilde{F}_R(\lambda))}{\lambda E[Y] (1-\widetilde{F}_Y(\lambda)\widetilde{F}_R(\lambda))},
\label{eq:Pae}
\end{equation}
where $\widetilde{F}_Z(.)$, for an arbitrary distribution function $F_Z(.)$, is given by:
\begin{equation}
\widetilde{F}_Z(\lambda)=\int_{0}^{\infty}{e^{-\lambda t}dF_Z(t)}.
\label{eq:tilda-function}
\end{equation}
For $Z=Y$ or $R$, this function gives the probability that the length of the operating or recovery period, respectively, is less than a packet inter-arrival time. Note that $P_{ae}$ is a conditional probability, conditioned on the fact that the system is empty. In general, $P_{ae}$ is only a function of the  moments of $Y$ and $R$. We therefore assume in the following that there is no correlation between $P_{ae}$ and $Y$ and $R$.  

The first moment of $X$ can be obtained by solving the following equation: 
\begin{equation}
E[X]=\rho E[X_b] + (1-\rho)E[X_e],
\label{eq:How-to-findEX}
\end{equation}
where $\rho=\lambda E[X]$ is the probability of the system being not empty.
The second moment of $X$ can be found equal to:
\begin{equation}
E[X^{2}]=\rho E[X^2_b] + (1-\rho)E[X^{2}_e].
\label{eq:Xe}
\end{equation}
%
We will now find the first two moments of $X_a$, $X_u$ and $X_b$.

\subsubsection{Arrival to an empty-available system}
For packets of case '$a$', as illustrated in Fig. \ref{fig-queue-analyze-Empty}.1, $X_a$, can be given by: 
\begin{equation}
\small
X_a=\begin {cases}
T,& Y_{1a}\geq T\\
Y_{1a}+R_1+Y_2+R_2+...+Y_K+R_K+T_R,& \mbox{Otherwise,}
\end{cases}
\label{eq:Xa}
\normalsize
\end{equation}
where $Y_{1a}$ is the random remaining time of the first operating period until the next interruption, $T_R$ is the transmission time of the last part of the packet, $K$ is the number of operating periods required to transmit the entire packet and
\begin{equation}
T=Y_{1a}+...+Y_K+T_R.
\label{eq:renewal-condition}
\end{equation}

If we consider the operating periods $\{Y_{1a},Y_2,\dots,Y_K\}$ as a renewal process, $K$ is the number of renewals of $Y$ during the real transmission time of a packet and  its distribution can be found from the renewal theory results \cite{cox62,ross06}. The first two moments of the number of renewals during $(0,t]$ composed of instances of $Y_{1a}$ and $Y$ are given by:
\begin{align}
m_a(t) & = \mathcal{L}^{-1}\left\{\frac{\widehat{Y}_{1a}(s)}{s(1-\widehat{Y}(s))}\right\}  \label{eq:laplace-of-diff-ma}, \\
m_a^2(t) & = \mathcal{L}^{-1}\left\{\frac{\widehat{Y}_{1a}(s)(1+\widehat{Y}(s))}{s\left(1-\widehat{Y}(s)\right)^2}\right\} \label{eq:laplace-of-diff-m2a}.
\end{align}
The moments of $K_a$ are then given by:
\begin{equation}
E[K_a]=\int_{0}^{\infty}{E[K_a|T=t] f_T(t) dt}=\int_{0}^{\infty}{m_a(t) f_T(t) dt}.
\label{eq:}
\end{equation}
 
We can rewrite (\ref{eq:Xa}) as:
\begin{equation}
X_a=T+\sum_{k=1}^KR_k.
\end{equation}
We then obtain the first moment of $X_a$ as:
\begin{equation}
E[X_a]=E[T]+E[K_a]E[R].
\label{eq:avg-xa}
\end{equation}
For the second moment, we use the fact that $K_a$ is independent of the recovery process, but not the service time, and that the variance of the random sum $\sum_{k=1}^KR_k$ is equal to $E[K]Var(R)+(E[R])^2Var(K)$ to obtain that  
\begin{align}
E[X^2_{a}]&=E[T^2]+2E[TK_a]{E[R]}+E[K_a](E[R^2]-(E[R])^2)+(E[R])^2{E[K_a^2]},
\label{eq:avg-x2a}
\end{align}
where
\begin{align}
E[TK_a]&=\int_{0}^{\infty}{E[TK_a|(T=t)] f_T(t)dt}=\int_{0}^{\infty}{t m_a(t) f_T(t)dt}.
\label{eq:cor-of-T-Ka}
\end{align}
\subsubsection{Arrival to a busy system}
Based on the distribution of the first operating period $Y_{1b}$ (see Fig. \ref{fig-queue-analyze-Busy}),
we can find the first two moments of $K_b$, the number of renewals for the case '$b$', and $X_b$ , as we did for $K_a$ and $X_a$ in the previous results.

\subsubsection{Arrival to an empty-unavailable system}
\label{app:stats}
For the case '$u$', as illustrated in Fig. \ref{fig-queue-analyze-Empty}.2, the completion time of the user is started within a recovery period and we have: 
\begin{equation}
X_u=R_{r}+Y_{1}+R_2+...+Y_K+R_{K+1}+T_R = R_{r} +X_u^*.
\label{eq:X_u}
\end{equation}

The moments of $X_u^*$ can be found as for $X_a$ by replacing $Y_{1a}$ with $Y_1$. $R_{r}$, the remaining time in the first recovery period, can be written as:
\begin{equation}
R_{r}=R-\mathbb{A}|(R>\mathbb{A}),
\label{eq:R1u}
\end{equation}
where $\mathbb{A}|(R>\mathbb{A})$ is the inter-arrival time conditioned on the fact that it should be less than $R$. The moments of $R_r$ based on $R$ and $\mathbb{A}$ are found as follows. 

As above, we encounter several times throughout the paper the random variables $Z=V|(V<U)$ and $Q=V-U|(V>U)$, for any two arbitrary random variables $V$ and $U$. We derive here the statistics of these two random variables. We have: 
\begin{equation}
f_{Z}(t)= \frac{Pr(U>t)f_{V}(t)}{Pr(U>V)} = \frac{(1-F_U(t))f_{V}(t)}{Pr(U>V)}. 
\label{eq:fY-X}
\end{equation}
When $U$ is exponentially distributed with parameter $\alpha$, we then have: 
\begin{equation}
Pr(V<U)=\int_{0}^{\infty}{e^{-\alpha t}f_V(t)}=\widehat{V}(\alpha),
\label{eq:PrX-Y-Exp}
\end{equation}
\begin{equation}
f_{Z}(t)= \frac{e^{-\alpha t} f_{V}(t)}{\widehat{V}(\alpha)}.  
\label{eq:fY-X-Exp}
\end{equation}
In this case, $E[Z]$ and $E[Z^2]$ can respectively be given by:
\begin{equation}
E[Z]= \frac{-d/d\alpha \widehat{V}(\alpha)}{\widehat{V}(\alpha)},  
\label{eq:EX-Y-Exp}
\end{equation}
\begin{equation}
E[Z^2]= \frac{d2/d\alpha^2 \widehat{V}(\alpha)}{\widehat{V}(\alpha)}.  
\label{eq:E2X-Y-Exp}
\end{equation}
For the second random variable, $Q$, we still assume that $U$ is exponentially distributed with parameter $\alpha$. Then, after some algebra manipulations (details can be found in \cite[Lemma2]{federgruen86} or in \cite{takagi91}), we obtain: 
\begin{equation}
E[Q]= \frac{E[V]}{1-\widehat{V}(\alpha)} - \frac{1}{\alpha},  
\label{eq:EQ}
\end{equation}
and 
\begin{equation}
E[Q^2]= \frac{E[V^2]-2\frac{E[V]}{\alpha}}{1-\widehat{V}(\alpha)} +  \frac{2}{\alpha^2}.  
\label{eq:E2Q}
\end{equation} 

Based on the moments of $R_{r}=R-\mathbb{A}|(R>\mathbb{A})$, the completion time can thus be found as:
\begin{align}
E[X_u]&=E[R_{r}]+E[X_u^*], \nonumber \\
E[X^2_u]&=E[R^2_{r}]+E[X^{*2}_u]+2E[R_{r}]E[X_u^*].
\label{eq:X_u}
\end{align}


\subsection{Queue performance metrics}
We can use the same approach used for M/G/1 queues to derive the waiting time for our system. When a packet arrives, it waits for the remaining completion time of the packet in service (if any), and then the completion time of all packets in the queue. For the packets which are in the queue, the completion time is always distributed with $X_b$ (they are queued, so they have not arrived to an empty system). However for the packet which is initially in service, the general completion time should be used because no knowledge is available to know whether this packet has been of case '$a$', '$b$' or '$u$'. We thus have:
\begin{equation}
E[W] = \frac{\lambda E[X^{2}]}{2(1-\lambda E[X_b])}.
\label{eq:waiting-by-fred}
\end{equation}
The average system time is given by 
$E[D] = E[W] + E[X]$.

\subsection{Busy periods}
\label{subsec:busy-periods}
Similarly to an M/G/1 queue without interruption \cite{takagi91}, we can find the busy periods' distribution for our queue with interruption. This result will be useful to analyze the priority disciplines. We know that the first completion time in a busy period is an instance of $X_e$. However, for other busy periods which are initiated during $X_e$, the busy period is started with an instance of $X_b$ because the packets enter a non-empty system. Therefore, we can find the LST of the busy periods as:
\begin{equation}
\widehat{B}(s)=\widehat{X}_e(s+\lambda-\lambda \widehat{B}_b(s)),
\label{eq:busypr-B}
\end{equation}
where $\widehat{B}_b(s)$ is the LST of the busy periods which are initiated during $X_e$ with an instance of $X_b$. $\widehat{B}_b(s)$ itself can be found from the following equation:
\begin{equation}
\widehat{B}_b(s)=\widehat{X}_b(s+\lambda-\lambda \widehat{B}_b(s)).
\label{eq:busypr-Bb}
\end{equation}
From the equation above, we can find the first and the second moments of $B_b(t)$ 
as:
\begin{equation}
E[B_b]=\frac{E[X_b]}{1-\lambda E[X_b]} \mbox{ and } E[B_b^2]=\frac{E[X_b^2]}{(1-\lambda E[X_b])^3}.
\label{eq:}
\end{equation}  
The first and the second moments of the general busy periods are then given by:
\begin{equation}
E[B]=\frac{E[X_e]}{1-\lambda E[X_b]},
\label{eq:busypr-E}
\end{equation}
\begin{equation}
E[B^2]=\lambda{E[B_b^2]}{E[X_e]} + (1+\lambda E[B_b])^2 E[{X_e}^2].
\label{eq:busypr-E2}
\end{equation}


\subsection{Alternative model}
\label{sec:alt_model}

An alternative model is to consider the start of the real service as the start of the completion time. We introduce this model since it will be useful to analyze some of the priority schemes. This alternative model does not affect the completion time for arrivals in empty available and busy systems (cases '$a$' and '$b$'), but for an arrival in an empty unavailable system, the remaining time of the recovery period $R_r$ is considered as waiting time. The completion time is given by $X_u^*$ (see~(\ref{eq:X_u})). The first two moments of $X^*$, the overall completion time for this alternative model, can then be found using the same approach as for $X$. Since the system time for both models must be the same we then have that the average waiting time for this model is given by:
\begin{equation}
E[W^*]=E[W]+E[X]-E[X^*].
\label{eq::waitingtime-case2}
\end{equation}
 
\subsection{Approximate and exponential operating periods}
\label{subsec:Exp-Av-periods}

As discussed in \cite{federgruen86}, in general it is very complex to find the exact distribution of $Y_{1a}$ and $Y_{1b}$, since they depend on the time when a packet arrives or a packet service has terminated. An approximation for $Y_{1a}$  and $Y_{1b}$ is to assume that they may be started uniformly during an operating period \cite{azarfar12e}, which is sometimes called \emph{random modification} of $Y$ \cite{avi-itzhak63} or \emph{equilibrium excess distribution} \cite{federgruen86}.

For the special, yet important, case that the operating periods are distributed with an exponential distribution $Y$ with parameter $\alpha$ (i.e., $F_{Y}=1-e^{-{\alpha}t}$), from the memoryless property we have that $Y_{1a}=Y_{1b}=Y$ and $K_a=K_u=K_b$. Using (\ref{eq:laplace-of-diff-ma}) and (\ref{eq:laplace-of-diff-m2a}), we have that $m(t)=\alpha t$ and $m^2(t)=\alpha^2 t^2+\alpha t$. It is then straightforward to derive the moments of the completion time as:
\begin{align}
E[X_{a,b,u}]&=E[T] (1+\alpha E[R]), \\
\label{eq:avg-xa-Exp}
E[TK_a]& =\alpha E[T^2],\\
\label{eq:ETKa}
E[X^2_{b}]& =E[T^2](1+\alpha E[R])^2+ \alpha E[T]E[R^2], \\
\label{eq:avg-x2a-Exp}
E[X^{2}_{u}] &=E[X^2_{a}]+E[R_{r}^2]+2E[X_{a}]E[R_{r}].
\end{align}


For exponentially distributed operating periods,  we can further model our queue as a queue with an initial setup time~\cite{takagi91} to find the waiting time. The initial setup time $S$ for a packet which initiates the busy period is $R_r$ with probability ($1-P_{ae}$) and zero otherwise. Thus, we can find the moments of $S$ based on the moments of $Y$ and $R$: $E[S]=(1-P_{ae})E[R_r]$ and $E[S^2]=(1-P_{ae})E[R^2_r]$. From \cite[(2.44a)]{takagi91}, we then have:
\begin{align}
E[D]&=E[X_b]+\frac{\lambda E[X_b^2]}{2(1-\lambda E[X_b])}+\frac{2E[S]+\lambda E[S^2]}{2(1+\lambda E[S])} \nonumber \\
&=E[X_b]+\frac{\lambda E[X_b^2]}{2(1-\lambda E[X_b])}+\frac{E[R^2]}{2(E[Y]+E[R])}.
\label{eq:eq:waiting-with-vacation2}
\end{align}
The steady-state probability of the system being empty, $P_0$, can be given by: 
\begin{equation}
P_0=\frac{E[I]}{E[I]+E[B_s]} = \frac{1-\lambda E[X_b]}{1+\lambda E[S]},
\label{eq:P0-original}
\end{equation}
where $E[I]=\frac{1}{\lambda}$ is the average of idle periods (no packet in the system), and $E[B_s]$ is the average of busy periods initiated by $S+X_b$ which can be found from Section \ref{subsec:busy-periods}.

\subsection{Case Study: Comparison Between Switching and Buffering OSA Strategies}
\label{subsec:case-study}
We now present a case study to validate the theoretical analysis and to discuss how it can be used to gain insight on the performance of OSA networks. 
In this case study, we compare two common OSA strategies which, following the detection of primary users activity on the operating channel, either \emph{switch} to a new channel or \emph{buffer} packets while waiting for the primary users to release the channel~\cite{park11,lai11}.

It is assumed that there is a large set of similar channels with exponentially distributed availability ($I$) and unavailability periods ($U$). For both OSA policies, we have $Y=I$ and for the buffering policy $R=U$ \cite{azarfar12e}. For the OSA switching policy, we use the common random sensing model in which the channels are sensed successively in a random order until an available channel is found. The interruption time $R$ is thus geometrically distributed with a success probability $\frac{E[I]}{E[I]+E[U]}$ for each time slot of $\tau$ ($\tau$ is the amount of time required to switch to and sense a channel). In the theoretical model, $R$ is approximated by an exponential distribution with an average length $E[R]=\frac{\tau(E[I]+E[U])}{E[I]}$.
\begin{figure}%
\includegraphics[scale=0.65]{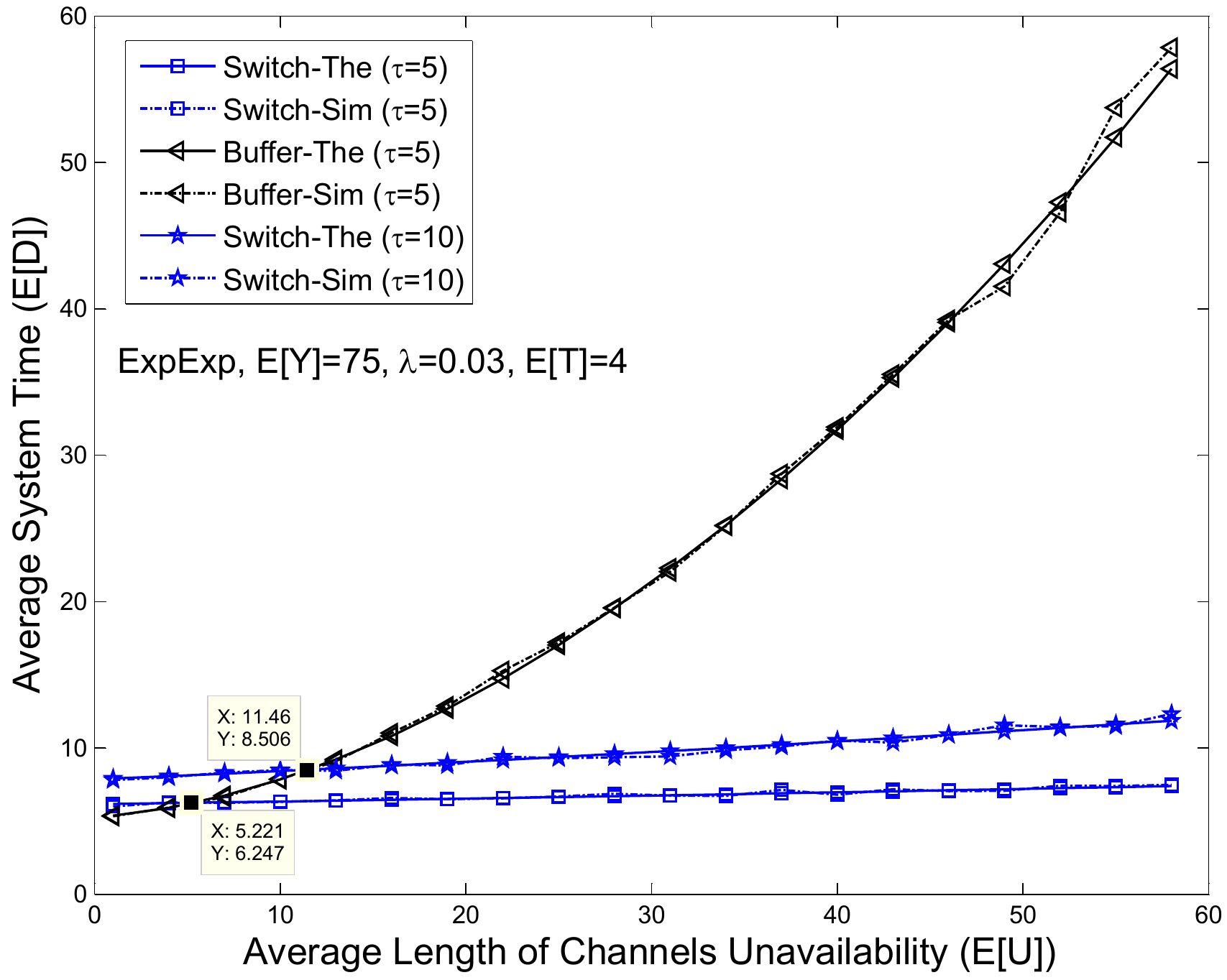}%
\caption{Decision on employing a buffering or switching policy fulfilled with analytical queueing results.}%
\label{Fig-Buff-Switch-Final}%
\end{figure}

Figure \ref{Fig-Buff-Switch-Final} compares the packet sojourn time ($E[D]$), for the system parameters indicated in the figure, of these two models obtained with exact Monte-Carlo simulations and using the theoretical result (\ref{eq:eq:waiting-with-vacation2}). First, the presented results confirm the accuracy of the theoretical model and its applicability to different OSA strategies. We can also observe that, as can be expected, the threshold point for the average channel unavailability length $E[U]$ where the switching policy becomes preferable over the buffering policy increases from 5.2 to 11.4 units of time when $\tau$ increases from 5 to 10 units of time. Note that this threshold is not simply given by the value $E[R]$ where the average interruption time for both policies are equal, but is obtained by finding the value of $E[R]$ where (\ref{eq:eq:waiting-with-vacation2}) is the same for both OSA policies. Based on the knowledge of the CR users sensing and switching time, and the estimated values of the average channel availability and interruption period lengths~\cite{gab13}, the OSA network can therefore use (\ref{eq:eq:waiting-with-vacation2}) to optimally decide between the switching and buffering policies to minimize the packets sojourn time.
In the remainder of the paper, we will derive similar relationships that can be used to analyze and optimally control on OSA network with multiple classes of traffic with priority queueing differentiated services. 

\section{Priority Queueing}
\label{sec:pr-queueing}

We can now tackle the analysis of the four priority queueing disciplines for the general queueing model with $N$ classes of CR traffic. Let $\rho_b=\sum_{j=1}^{N}{\rho_{b,j}}=\sum_{j=1}^{N}{\lambda_j E[X_{b,j}]}$ and $\mathbb{A} = \min \{ A_1,...,A_{N} \} \rightarrow \mathbb{A} \sim EXP(\lambda)$.
We also use the notation $P_{ae}(\lambda)$ and $R_r(\mathbb{A})$ to highlight that $P_{ae}$ and $R_r$ in (\ref{eq:Pae}) and (\ref{eq:R1u}), respectively, should be calculated with combined $\lambda$ and $\mathbb{A}$. 
For the non-preemptive and preemptive priority queueing disciplines presented in Section~\ref{sec:non_preempt} and~\ref{sec:preempt}, respectively, results for a general distribution for the operating periods are presented. For the exceptional non-preemptive and preemption in case of failure service disciplines, introduced in this paper and presented in Section~\ref{sec:except_non_preempt} and~\ref{sec:preempt_failure} respectively, we only analyze the case of exponential operating periods due to the analytical complexity of those schemes without the assumption of memoryless operating periods. We also present in Section~\ref{app:alt_approach} an alternative approach to analyze the preemptive and non-preemptive disciplines for exponential operating periods.

\subsection{{Non-preemptive}}
\label{sec:non_preempt}

Since the packet service can not be preempted in this scheme, the completion time of any packet for the three cases ('$a$','$b$' and '$u$') will be the same as for the single traffic queue. The moments of the general completion time $X_i$ for class of traffic $i$, $i=1,\dots,N$, can then be found by solving the system of $N$ equations and $N$ unknowns obtained from (\ref{eq:How-to-findEX}) for the $N$ classes of traffic where $\rho$ is replaced by $\sum_{j=1}^{N}{\lambda_j E[X_j]}$.
Then, similar to an M/G/1 queue \cite{bertsekas92}, we have: 
\begin{equation}
E[W_i]=\frac{E[J]}{(1-\sum_{j=1}^{i}{\rho_{b,j}})(1-\sum_{j=1}^{i-1}{\rho_{b,j}})}.
\label{eq:waiting-resume-nonpr}
\end{equation}
where $J$ is the remaining completion time of the packet in service and is given by:
\begin{equation}
E[J]=\sum_{j=1}^{N}{\frac{\lambda_j}{2}E[X_{j}^{2}]}.
\label{eq:}
\end{equation}
Note that since no knowledge is available about the packet in service, the general completion time is used. However, the denominator represents the completion time of the queued packets which is $X_{b,j}$ for class $j$. 

When $Y$ is exponentially distributed, we have that the completion time for the three cases ('$a$','$b$' and '$u$') has the same distribution in the alternative model presented in Section~\ref{sec:alt_model}. Using the same approach as for (\ref{eq:eq:waiting-with-vacation2}), a closed-form relation can be obtained for the system time
by using a queue model with an exceptional completion time $X_e$ for the first packet which initiates a busy period \cite{takagi91}. $X_e$ is given by:
\begin{align}
X_e&=\sum_{i=1}^{N}{\frac{\lambda_i}{\lambda} X_{e,i}}=\sum_{i=1}^{N}{\frac{\lambda_i}{\lambda} \left[P_{ae}(\lambda)X_{b,i}+(1-P_{ae}(\lambda))(X_{b,i}+R_r(A))\right]}.
\label{eq:}
\end{align}
It is straightforward to find the first two moments of $X_{e,i}$ and $X_{e}$. We then obtain \cite{takagi91}:
\begin{align}
E[D_i]&=\frac{(1-{\rho}_b)(E[X_{e,i}]+E[{X_{b,i}}]) + \lambda E[X_{e,i}]E[{X_{b,i}}]}{1+\lambda E[X_e]-{\rho_b}} \nonumber \\
&+\frac{\lambda[(1-{\rho_b})E[X_e^2]+E[X_e](\sum_{j=1}^{N}{{{\lambda}_j}E[({X_{b,j}})^2]})]}{2(1+\lambda E[X_e]-\rho_b){(1-\sum_{j=1}^{i}{\rho_{b,j}})(1-\sum_{j=1}^{i-1}{\rho_{b,j}})}}.
\label{eq:waiting-resume-nonpr-Ex}
\end{align}
The first term represents the average completion time and the second term, the average waiting time. 
Similar to (\ref{eq:P0-original}), $P_0$, the steady-state probability of the system being empty, is given for this queue by:
\begin{equation}
P_0= \frac{1-\rho_b}{1-\rho_b+\lambda E[X_e]}.
\label{eq:P0-nonpr}
\end{equation}

\subsection{{Exceptional non-preemptive}}
\label{sec:except_non_preempt}

In this scheme, a packet which arrives in an empty unavailable system (case '$u$') can be preempted at the end of the arrival recovery period by a higher priority packet which also arrives in the same recovery period. However, this is in fact a non-preemptive discipline for the alternative model presented in Section~\ref{sec:alt_model} since in this model a packet which arrives in an empty unavailable system does not start the service, but is queued waiting to obtain the server which will be given to the queued packet with the highest priority.
We thus have a non-preemptive queue with initial setup time~\cite{takagi91}. Using the same approach as in Section~\ref{subsec:Exp-Av-periods}, we obtain that:
\begin{align}
E[D_i]&=E[X_{b,i}]+\frac{\sum_{j=1}^{N}{\lambda}_j E[({X_{b,j}})^2]}{2{(1-\sum_{j=1}^{i}{\rho_{b,j}})(1-\sum_{j=1}^{i-1}{\rho_{b,j}})}} \nonumber \\
&+\frac{(1-{\rho_b})(\lambda E[S^2]+2E[S])}{2(1+\lambda E[S]-\rho)(1-\sum_{j=1}^{i}{\rho_{b,j}})(1-\sum_{j=1}^{i-1}{\rho_{b,j}})}.
\label{eq:waiting-Exnonpr-Exp}
\end{align}
The probability of the system being empty is equal to: 
\begin{equation}
P_0=\frac{1-\rho_b}{1+\lambda E[S]}.
\label{eq:P0-ENP}
\end{equation}

\subsection{{Preemptive}}
\label{sec:preempt}

In this scheme, the highest class is not affected by the other classes of traffic. Its completion time and system time can thus directly be found using the results presented in Section~\ref{sec:queue-analysis}. Let us now analyze the performance of the low priority class for a two priority class system.

\begin{figure}%
\includegraphics[scale=0.65]{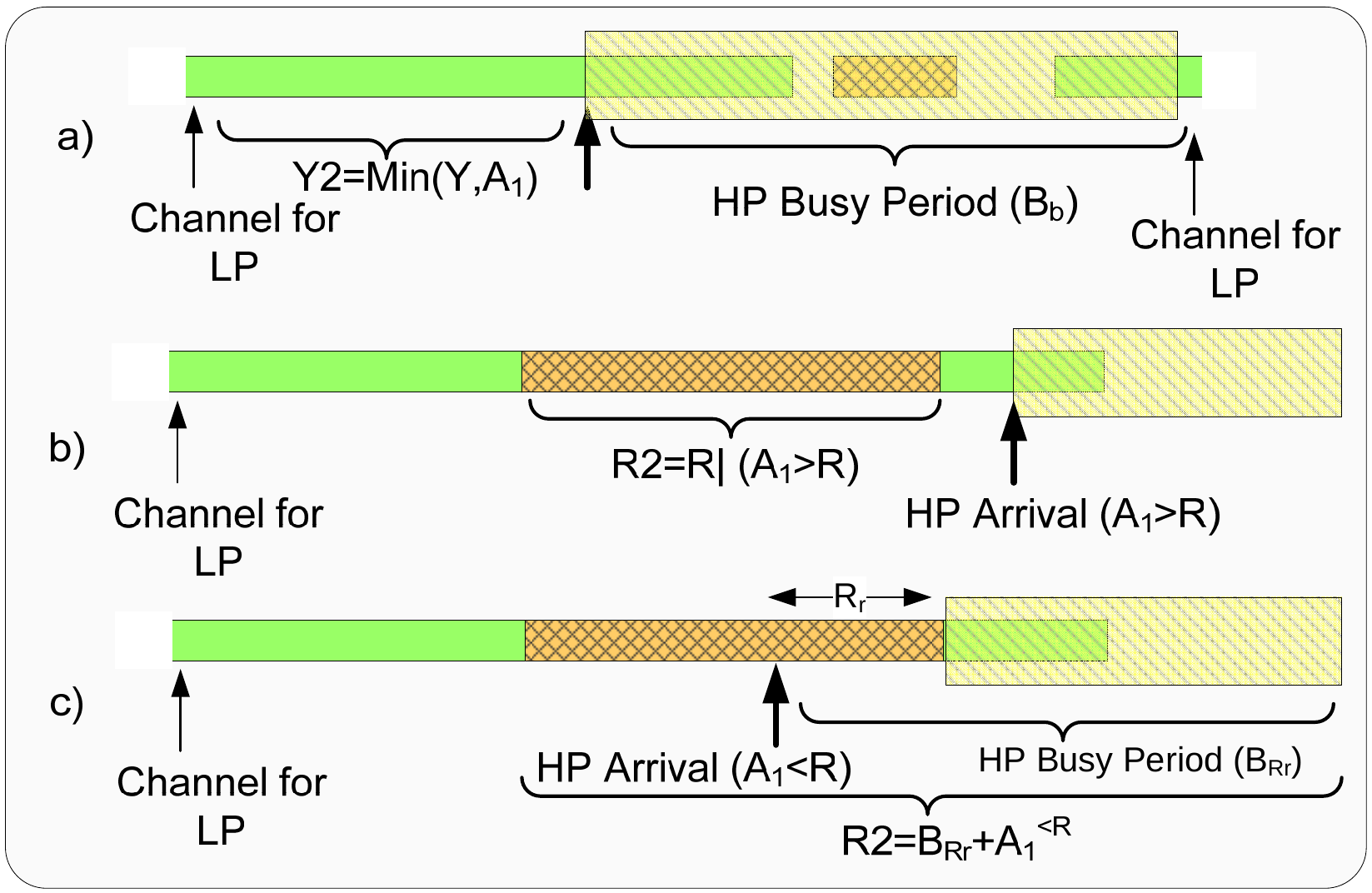}%
\caption{Operating and interruption periods ($Y_2$ and $R_2$) from the perspective of LP packets.}%
\label{fig-Y2R2Dist}%
\end{figure}

To solve this system, let us find the distribution of $Y_2$ and $R_2$, respectively the operating and interruption periods, from the perspective of the low priority (LP) packets\footnote{They are here two new random variables. Not to be mistaken with identical instances of $Y$ and $R$ in previous sections.}. The system is unavailable for LP traffic both due to the activity of high priority (HP) users and due to channel interruptions. As illustrated in Fig. \ref{fig-Y2R2Dist}.a, $Y_2$ is the minimum between the time to the next interruption and the arrival of an HP packet: any one which arrives sooner initiates an interruption period for LP packets. We thus have:
\begin{equation}
Y_2=\min (Y,A_1)  \rightarrow  1-F_{Y_2}(t) = (1-F_{Y}(t)) (1-F_{A_1}(t)).
\label{eq:Dist-Y2-base}
\end{equation}
When $Y$ is exponentially distributed with parameter $\alpha$, the distribution of $Y_2$ can be given by: 
\begin{equation}
F_{Y_2}(t) = 1- e^{-(\lambda_1+\alpha)t}.
\label{eq:Dist-Y2-base}
\end{equation}
To calculate $R_2$, we have to distinguish between the events that caused the period of interruption.
If, as illustrated in Fig. \ref{fig-Y2R2Dist}.a, a high priority (HP) packet arrived and preempted the low priority (LP) traffic, the length of $R_2$ is equal to one busy period of HP packets which is distributed according to $B_{b,1}$. On the other hand, if the channel interruption caused the unavailability, the two cases  shown in Fig. \ref{fig-Y2R2Dist}.b and \ref{fig-Y2R2Dist}.c may happen. First, if no HP packet arrives during $R$, the length of $R_2$ is equal to $R|R<A_1$. If an HP packet arrives during $R$, $R_2$ will be $A_1|(R>A_1)$, the interruption period until the HP packet arrival, extended with an HP busy period $B_{R_r}$. $B_{R_r}$ can be found from (\ref{eq:busypr-B}), replacing $X_e$ by $X_{b,1}+R_r$ where $R_{r}=R-A_1|(R>A_1)$ is the remaining time of the server interruption after the HP packet arrival and $X_b$ by $X_{b,1}$. We then have:
\begin{equation}
R_2=\begin{cases}
B_{b,1} & Pr(A_1<Y),\\ 
R|(R \leq A_1)& Pr(Y \leq A_1 {\&} R \leq A_1),\\
A_1|(R>A_1)+B_{R_r}& Pr(Y \leq A_1 {\&} R>A_1).
\end{cases}
\label{eq:}
\end{equation}
Please be aware that in order to simplify the notation, from now on, we use the notation $V^{<Z}$  to denote $V|(V<Z)$ for any two random variables $V$ and $Z$.
The probability of an HP arrival during $R$  can be given by:
\begin{equation}
P_{A_1<R}=\int_{0}^{\infty}{(1-e^{-\lambda_1 r})dF_R(r)}.
\end{equation}
For exponential $Y$, we have:
\begin{equation}
Pr(Y \leq A_1)= \frac{\alpha}{\alpha + \lambda_1},
\label{eq:}
\end{equation}
and then
\begin{align}
E[R_2]&=\frac{\lambda_1}{\alpha + \lambda_1} E[B_b]+\frac{\alpha}{\alpha+\lambda_1}\left[(1-P_{A_1<R})\frac{-d/d\lambda_1 \widehat{R}(\lambda_1)}{\widehat{R}(\lambda_1)}+P_{A_1<R}(E[A_1^{<R}]+E[B_{R_r}])\right].
\end{align}
Using (\ref{eq:EQ}) for $E[R_r]$ and (\ref{eq:busypr-E}) for busy periods, we have:
\begin{equation}
E[B_{R_r}]=\frac{E[X_{b,1}]+E[R_r]}{1-\lambda_1E[X_{b,1}]}= \\
\frac{E[X_{b,1}]+\frac{E[R]}{1-\widehat{R}(\lambda_1)}-\frac{1}{\lambda_1}}{1-\lambda_1E[X_{b,1}]}.
\label{eq:}
\end{equation}

The second moment of $R_2$ can be computed similarly, where the second moment of the busy periods can be found from (\ref{eq:busypr-E2}) and the second moment of $A_1^{<R}$ and $R^{<A_1}$ can be derived from (\ref{eq:E2X-Y-Exp}) in Section~\ref{app:stats}. It should be taken into account that $B_{R_r}$ and $A_1^{<R}$ are correlated, so $E[B_{R_r} A_1^{<R}]$ should be calculated separately, using, for instance, the same approach as in (\ref{eq:cor-of-T-Ka}). 


From the equations above, one can find the moments of the operating and interruption periods from the perspective of LP packets ($Y_2$ and $R_2$). Then, we return to the original M/G/1 queue with interruptions and replace $Y$ and $R$ in (\ref{eq:eq:waiting-with-vacation2}) with $Y_2$ and $R_2$, respectively, to find the performance metrics of the LP packets. 

Note that it is not easy to extend the proposed approach for more than two classes of CR traffic; however, it can be used to find a bound for the performance of aggregated low priority traffic (combination of all low priority classes).

\subsection{{Preemption in case of failure}}
\label{sec:preempt_failure}

In this priority queueing model, high priority packets can only preempt the service from a low priority packet if an interruption occurs. When the service is resumed after an interruption, the priority is given first to high priority packets (HP). In other words, if an HP packet arrives while a low priority (LP) packet is in service, the HP service is started either after the end of the LP service or after an interruption, any one which occurs sooner. As expected, for any class of traffic the performance metrics for this scheme lies between the non-preemptive and preemptive schemes.  
The completion time of HP packets is not affected by this service discipline and can be found from the original single traffic queue. However, the waiting time of the HP packet is affected since it must wait until the end of the LP packet transmission or an interruption before starting its service. In the following, we first analyze the completing time of LP packets with this service discipline and then study the HP and LP waiting time. Finally, we discuss the special case where the service time of low priority packets is exponentially distributed. 

\subsubsection{Completion time of the LP packets} 
To find the completion time of the LP packets, we follow a similar approach as the one used for the preemptive scheme. Due to the memoryless operating periods distribution, from the LP packets perspective, the operating period distribution is not affected by the preemptive in case of failure service discipline and we have $Y_2=Y$. On the other hand, 
$R_2$, the length of the interruption period from the LP users perspective, is a function of the remaining service time of the LP packet at the HP packet arrival time. That is, the longer the remaining service time until the next interruption, the more HP packets can arrive and thus their busy period will get longer. However, unless the service time is memoryless (this special case is discussed in Section~\ref{sec:preempt_failure_exp}), the remaining service time is not the same for each interruption $R_2$. Therefore, the completion time can not be modeled as a renewal process because the instances of $R_2$ are not identical. 
We will thus provide approximations for the moments of $X_2$ (or $X^*_2$) for two extreme cases: when the operating periods are much larger than the service time of type-2 (LP) packets ($Y >> T_2$) (\emph{large scenarios}) and when it is smaller ($Y < T_2$) (\emph{small scenarios}).  

For $Y >> T_2$, it can be assumed that the service of an LP packet is finished in at most two type-2 operating periods. This assumption is a trade-off between accuracy and complexity and it is  equivalent to assuming at most one instance of $R_2$ interruption during the completion time. The completion time will be found based on the alternate model (Section \ref{sec:alt_model}). The expectation of $X^*_2$ is then given by:
\begin{equation}
E[X^*_2] \approx \begin{cases}
E[T_2]& Y \geq T_2 \\
E[T_2]+E[R_2]& Y<T_2.
\end{cases}
\label{eq:}
\end{equation}
The length of $R_2$ depends on the arrival of an HP packet and its arrival time. 
$R_2$ can be given by:
\begin{equation}
R_2=\begin{cases}
R^{<A_1}& \mbox{No HP arrival},\\
B_{C_r}-(Y^{<T_2}-A_1|(A_1<Y^{<T_2}))& \mbox{HP arr. in $Y^{<T_2}$},\\
A_1^{<R}+B_{R_r}& \mbox{HP arr. in $R$}.\\
\end{cases}
\label{eq:R2-appx1-newpr}
\end{equation}
where $C_r$ is the remaining time of the cycle (a cycle consists of an operating period $Y$ followed by a recovery period $R$) after the arrival of an HP packet, and $B_{C_r}$ represents the HP busy period which is initiated with $C_r+X_{b,1}$. However, $Y^{<T_2}-A_1$, the remaining time of the operating period $Y^{<T_2}$ after the HP packet arrival, should be excluded from $R_2$ since the HP packet does not immediately preempt the LP packet. In the third case, the HP arrival occurs in $R$. The busy period of HP packets thus starts with $R_{r}+X_{b,1}$ and the length of the total interruption is $A_1$ in addition to the HP busy period. $E[R_2]$ can thus be given by:
\begin{align}
E[R_2]&=(1-P_{A_1<C^{<T_2}}) \frac{-d/d\lambda_1 \widehat{R}(\lambda_1)}{\widehat{R}(\lambda_1)} +P_{A_1<C^{<T_2}}\nonumber\\
& \left[P_{ae}(\lambda_1)\Bigg(\frac{E[R]+E[Y^{<T_2}-A_1|A_1<Y^{<T_2}]+E[X_{b,1}]}{(1-\lambda_1E[X_{b,1}])} \right. \nonumber\\
& \left. -E[Y^{<T_2}-A_1|A_1<Y^{<T_2}]\Bigg)+(1-P_{ae}(\lambda_1))\Bigg(E[A_1^{<R}] + \frac{E[X_{b,1}]+E[R_r]}{(1-\lambda_1E[X_{b,1}])}\Bigg)\right],
\label{eq:eq:R2-appx2-newpr}
\end{align}
where $P_{A_1<C^{<T_2}}$ is the probability of an arrival in $C^{<T_2}=Y^{<T_2}+R$, and $P_{ae}$ is calculated for HP packets. The second moment of $R_2$ can be found similarly using the second moment of the busy periods and the relations provided in Section~\ref{app:stats}. However, the correlation of random variables $B_{R_r}$ and $A_1^{<R}$ should be taken into account. 

\begin{figure}%
\includegraphics[scale=0.65]{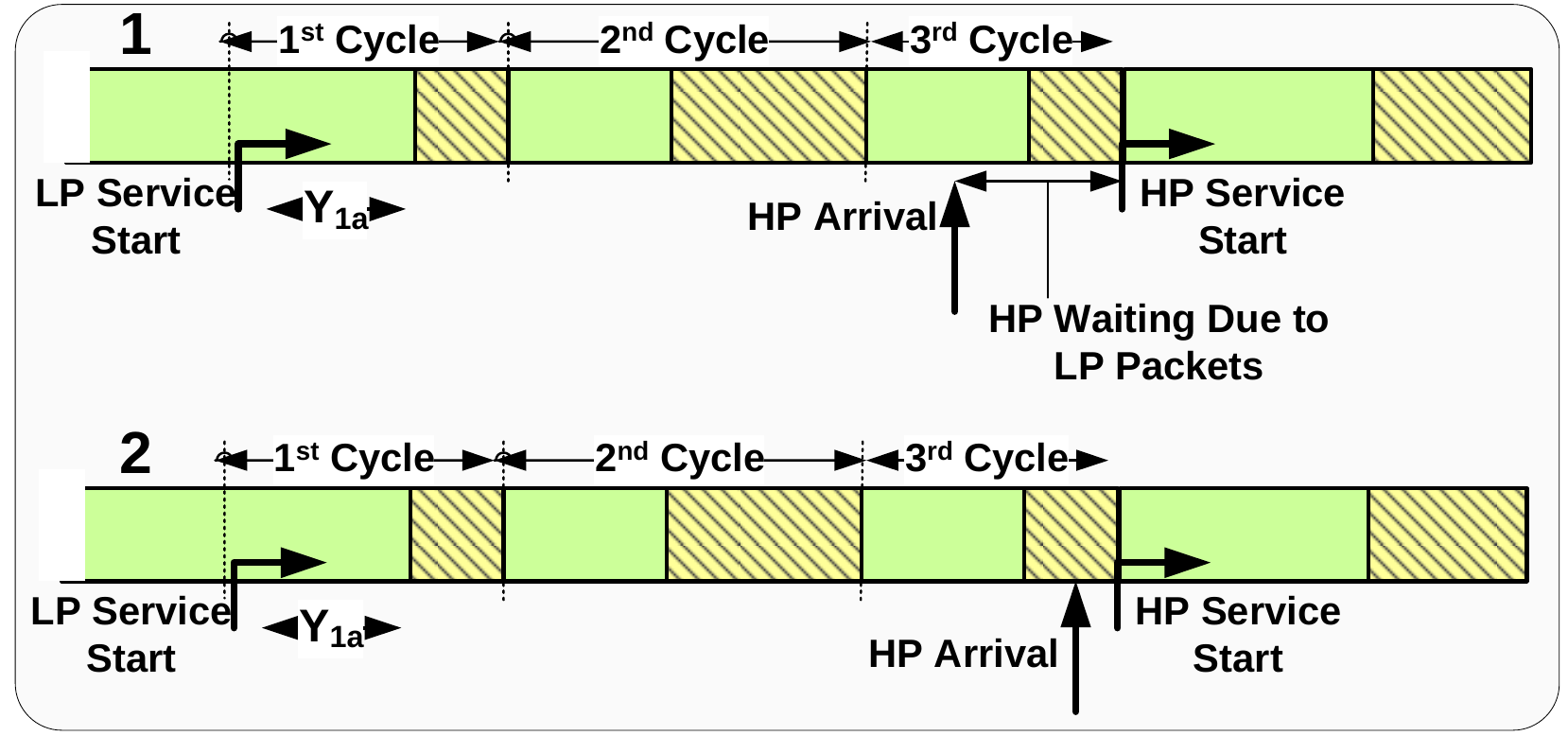}%
\caption{Cycles and holding periods for the LP packets in the discipline of preemptive in case of failure (FP). The LP packet can hold the channel for three cycles. However in a preemptive scheme (Pr) (not shown here), it can keep the channel for two complete cycles and releases the channel in the middle of the third cycle if there is an HP arrival (section 1). }
\label{fig-queue-cycles}%
\end{figure}

For the case where $Y < T_2$, we assume that the duration of HP busy periods is independent of the activity of LP packets. Therefore, the interruption periods from the perspective of LP packets have the same distribution and  a renewal process can be considered to analyze the completion time. When a LP packet starts its service, as illustrated in Fig. \ref{fig-queue-cycles}, it holds the channel (available or unavailable) for a cycle and if there is an HP arrival during the cycle, it releases the channel to HP packets at the end of the cycle. So, the probability of releasing the channel to HP traffic at the end of the cycle is given by:
\begin{equation}
P_{A_1<C}=Pr(\mbox{HP arrival in $C$}) 
=\int_{0}^{\infty}{(1-e^{-\lambda_1 c}) f_C(c)dc}.
\label{prob-of-an-arrival-in-c}
\end{equation}
For $R_2$, (\ref{eq:R2-appx1-newpr}) is still valid except that we eliminate the condition that $Y<T2$.

We then obtain:
\begin{align}
E[R_2]&=(1-P_{A_1<C}) \frac{-d/d\lambda_1 \widehat{R}(\lambda_1)}{\widehat{R}(\lambda_1)} \nonumber \\
&+P_{A_1<C}\left[P_{ae}(\lambda_1)\Bigg(\frac{E[R]+E[Y]}{(1-\lambda_1E[X_{b,1}])}-E[Y]\Bigg)+(1-P_{ae}(\lambda_1))\Bigg(E[A_1^{<R}] + \frac{E[X_{b,1}]+E[R_r]}{(1-\lambda_1E[X_{b,1}])}\Bigg)\right].
\label{eq:eq:R2-appx2-newpr}
\end{align}
The second moment can be found similarly. The moments of $R_2$ can then be substituted into the results for the single traffic queue  to find the moments of the completion time of LP packets $E[X_2]$ and $E[X_2^2]$.

\subsubsection{Waiting time}
\label{subsec:prf-Waiting-time}
For the high priority (HP) traffic, the waiting time of HP arrivals during $Y$ in a system empty of HP packets is affected by the presence of LP packets. For those packets, the new waiting time is zero if the LP queue is empty or, if the LP queue is not empty, the minimum of the remaining time of the arrival cycle and the remaining service time of the LP packet in service. The difficulty to compute the waiting time is thus the dependency of both types of traffic on each other. That is, while the waiting time of the HP packets is affected by the lower class, both the waiting time and the completion time of LP packets are affected by HP traffic. This obliges us to use approximations and bounds to find the waiting time of LP and HP packets. 


The waiting time of HP packets is upper bounded by an M/G/1 queue with vacation in (\ref{eq:eq:waiting-with-vacation2}) if we neglect the unknown part of the service that the LP has received so far and assume that the remaining LP service time is still $T_2$. That is, the initial setup time $S$ can be approximated as follows:
\begin{equation}
S=\begin{cases}
R_r(\lambda_1)& 1-P_{ae}(\lambda_1), \\
Y^{<T_2}+R&  Pr(Y \leq T_2)(1-P_0)P_{ae}(\lambda_1),\\
T_2^{<Y}& Pr(Y>T_2)(1-P_0) P_{ae}(\lambda_1),\\
0& P_0 P_{ae}(\lambda_1),
\end{cases}
\label{eq:W1-Estimation-Exp-NewPr-Apx}
\end{equation}
where $P_{ae}(\lambda_1)$ and $R_r(\lambda_1)$ only take HP packets into account, and $P_0$ is the probability of system being empty of any type of packet. Note that the correct probability to be used here instead of $1-P_0$ is $P_{L|NH}$, which is the probability that there are LP packets in the system given that it is empty of HP packets. However, this probability can not be found without any assumption on $T_2$'s distribution; therefore, we used $P_0$ as an approximation. This approximation results in a setup time, $S$, larger than its real value, which provides an upper bound in (\ref{eq:eq:waiting-with-vacation2-HP}) that can be sometimes looser than expected.  
$P_0$ is the same for all priority queueing models and is given in (\ref{eq:P0-nonpr}).
An obvious lower bound on the waiting time is given by assuming that HP packets always preempt LP packets (preemptive discipline). We can thus write:
\begin{equation}
\begin{split}
\frac{\lambda E[X_{b,1}^2]}{2(1-\lambda E[X_{b,1}])}+\frac{E[R^2]}{2(E[Y]+E[R])} \leq  E[W^*_1] < \frac{\lambda E[X_{b,1}^2]}{2(1-\lambda E[X_{b,1}])}+\frac{2E[S]+\lambda_1 E[S^2]}{2(1+\lambda E[S])}.
\end{split}
\label{eq:eq:waiting-with-vacation2-HP}
\end{equation}
We can then use those bounds on the HP waiting time to find corresponding bounds on the LP waiting time through the Conservation Law (CL) in a queue with multiple classes of traffic \cite{kleinrock75} which indicates that the quantities
\begin{equation}
\kappa = \lambda_1 E[T_1] E[W_1] + \lambda_2 E[T_2] E[W_2],
\label{eq:conservation-law}
\end{equation}
and  
\begin{equation}
\kappa^* = \lambda_1 E[T_1] E[W^*_1] + \lambda_2 E[T_2] E[W^*_2],
\label{eq:conservation-law}
\end{equation}
for alternative model, are constant for all priority service disciplines. $\kappa$ and $\kappa^*$ can thus be computed with the waiting time of LP and HP packets found for one of the previous priority disciplines.  

An alternative approach is to directly find the LP waiting time and then use the conservation law to obtain the HP waiting time. But, similarly to the HP waiting time, it is difficult to find an exact expression for the LP waiting time due to the strong interdependence between both types of traffic. We thus propose to compute bounds as follows.
Using the approximations for the first two moments of $X_2$ or $X^*_2$ found previously, the minimum waiting time of LP packets can be found using the {Pollaczek–-Khinchine} ({P-K}) relation. An upper bound for the LP packets' waiting time is naturally given by the waiting time in the preemptive discipline model. 

We thus have two upper and lower bounds for both the HP and LP waiting time. The tighter bounds can then be selected as the final lower and upper bound for both traffic categories.

\subsubsection{Exponentially distributed LP service time}
\label{sec:preempt_failure_exp}

The queue model for the HP traffic is an M/G/1 queue with vacations. Since for a memoryless exponentially distributed service time, the remaining parts of the service are identically distributed, it is possible to exactly express $S$, the initial setup time for the HP traffic, as follows:
\begin{equation}
S=\begin{cases}
R_r(\lambda_1)& 1-P_{ae}(\lambda_1), \\
Y^{<T_2}+R&  Pr(Y \leq T_2)P_{L|NH}P_{ae}(\lambda_1),\\
T_2^{<Y}& Pr(Y>T_2)P_{L|NH}P_{ae}(\lambda_1),\\
0& Otherwise, 
\end{cases}
\label{eq:W1-Estimation-Exp-NewPr}
\end{equation}
where $P_{ae}(\lambda_1)$ and $R_r(\lambda_1)$ only take HP packets into account. $E[S]$ is then given by:
\begin{align}
E[S]&=(1-P_{ae}(\lambda_1))E[R_r]+P_{L|NH}P_{ae}(\lambda_1)\left[(\frac{1}{\gamma_2+\alpha}+\frac{\alpha}{\alpha+\gamma_2} E[R])\right].
\label{eq:ES-NewPr-Exp}
\end{align}
$\alpha$ is the exponential parameter for $Y$ and $\gamma_2$ is the exponential parameter for $T_2$.
The unknown in the preceding equation is $P_{L|NH}$, which is the probability that there are LP packets in the system given that it is empty of HP packets, and can be expressed as:
\begin{equation}
P_{L|NH}= 1-\frac{P_0}{P_{0,1}},
\label{eq:PLNH-NewPr-Exp}
\end{equation}
where $P_{0,1}$, the probability that the system is empty of HP packets,  can be found from the original queue by substituting $E[S]$ with the one that was calculated in (\ref{eq:ES-NewPr-Exp}), and, as indicated previously, $P_0$ is the probability that the system is empty of all packets and is given in (\ref{eq:P0-nonpr}). We thus have two equations ((\ref{eq:ES-NewPr-Exp}) and (\ref{eq:PLNH-NewPr-Exp})) which can be used to find the two unknowns $E[S]$ and $P_{L|NH}$. The second moment of $S$ can then be found and the HP waiting time is given by (\ref{eq:eq:waiting-with-vacation2}).


The LP packets waiting time can be found easily from the conservation law and the previous result for the HP packets waiting time. To find the system time, we will now find an exact expression for the first two moments of the completion time of LP packets. The interruptions, from the LP packets point of view, are identically distributed due to the exponential LP service time distribution.
The same renewal process approach which was used as an approximation for $Y<T_2$ duration in (\ref{eq:eq:R2-appx2-newpr}) can thus be used exactly for this case. The only change is that $Y$ should be replaced with $Y^{<T_2}=Y|(Y<T_2)$.
We then have 
\begin{align}
E[R]&=(1-P_{A_1<Y|Y<T_2}-P_{A_1<R})E[R^{<{A_1}}] \nonumber \\
&+ P_{A_1<Y|Y<T_2}(E[B_{C_r}]-E[Y^{<T}])+P_{A_1<R} (E[B_{R_r}]+E[A_1^{<R}]),  
\end{align}
where 
\begin{align}
P_{A_1<Y|Y<T_2}&=\frac{\lambda_1}{\lambda_1+\alpha+\gamma}, \\
P_{A_1<R}&=(1-P_{A_1<Y|Y<T_2}){Pr(R<A_1)}, \\
E[B_{C_r}]&=\frac{E[R]+\frac{1}{\gamma_2+\alpha}+E[X_{b,1}]}{1-\lambda_1E[X_{b,1}]}, \\
E[B_{R_r}]&=\frac{E[R_{r}]+E[X_{b,1}]}{1-\lambda_1E[X_{b,1}]}.
\end{align}
Similarly, $E[R_2^2]$ and consequently $E[X_2]$ and $E[X^2_2]$ can be found.

It is worth noting that only the assumption of exponential service time for LP packets is required to find the results above. The service time of HP packets can be general. 

\section{Alternative approach for priority queueing analysis}
\label{app:alt_approach}
In this section, we propose an alternative approach to analyze the preemptive and non-preemptive priority queueing service disciplines in OSA networks. As interruptions have a preemptive behavior, we can model the interruptions as the highest priority type of traffic whose packet inter-arrival time is distributed with a random variable $Y$ and whose \emph{busy periods} are distributed with a random variable $R$. We can then have an estimate for the service time of these virtual highest priority packets whose  busy period models the interruptions. Closed-form relations can not be derived in general since for any distribution of $R$ and $Y$, a different formula for the busy periods and consequently for the service time exists. If $Y$ is exponentially distributed with parameter $\alpha$, we can assume an  M/G/1 queue for the interruptions.  
Note that, as discussed in the introduction, this \emph{alternative approach provides an approximation} because we know that the real distribution of the busy periods in an M/G/1 queue is a complicated function built on the Bessel function \cite{takagi91}, which can not exactly be matched to $R$. We discussed this alternative approach in this paper for completeness and as an extension of the analysis provided in~\cite{wang-l11,rashid07,laourine10} for multiple classes of traffic with preemptive and non-preemptive service disciplines.

Using this alternative approach, it is possible to find approximate results for the preemptive and non-preemptive service disciplines as follows.
We first find the first two moments of the service time of the virtual packets which form the interruptions from the distribution of the busy periods of a regular M/G/1 queue \cite{takagi91}: 
\begin{equation}
\widehat{R}(s)=\widehat{T}_v(s+\alpha-\alpha\widehat{R}(s)),
\label{eq:}
\end{equation}
\begin{equation}
E[T_v]=\frac{E[R]}{1+\alpha E[R]} \mbox { ,  } E[T^2_v]=E[R^2] (1-\alpha E[T_v])^3.
\label{eq:Moments-ofoTv}
\end{equation}
where $T_v$ stands for the service time of the virtual packets which form the interruptions.
As interruptions have a preemptive behavior, we can model the preemptive service discipline as a preemptive-resume queue with $N+1$ classes of traffic, the highest priority packets being the virtual packets.
The following extensions of the P-K formula for preemptive-resume schemes \cite{takagi91} can then be used to find the average waiting time of other classes of traffic:
\begin{equation}
\small
\begin{split}
E[W_i]= \frac{E[J_i]}{(1-\alpha{E[T_v]}-...-\lambda_{i-1} E[T_{i-1}])(1-\alpha{E[T_v]}-...-\lambda_i E[T_{i-1}])},
\label{eq:pk_ext}
\end{split}
\normalsize
\end{equation}
where $E[J_i]$ can be given by:
\begin{equation}
E[J_i]= \frac{1}{2} \alpha E[T_v^2]+ \sum_{j=1}^{i}{\frac{1}{2} \lambda_j(E[T_j^2])},
\label{eq:}
\end{equation}
and the moments of $T_v$ can be found from (\ref{eq:Moments-ofoTv}).
The system time is then given by:
\begin{equation}
E[D_i]=\frac{E[T_i]}{1-\rho_{i-1}-\dots-\rho_1-\rho_v}+E[W_i]
\end{equation}

For the non-preemptive service discipline, the queue can be modeled as a priority queue where the highest class of traffic (virtual) behaves preemptively, but other classes behave non-preemptively.  The extensions of the P-K formula given in (\ref{eq:pk_ext}) can be used to find the average waiting time of other classes of traffic where $E[J_i]=E[J]$ is the same for all priority classes and is equal to:
\begin{equation}
E[J]= \frac{1}{2} \alpha E[T_v^2]+ \sum_{j=1}^{N}{\frac{1}{2} \lambda_j(E[T_j^2])}.
\label{eq:}
\end{equation}
The system time can be written as: 
\begin{equation}
E[D_i]= \frac{E[T_i]}{1-\alpha{E[T_v]}}+E[W_i].
\label{eq:}
\end{equation}

\section{Simulation Results}
\label{sec:sim-analysis}
In this section, we validate the analytical results of the OSA networks priority queueing disciplines presented in this paper by comparing with system accurate Monte-Carlo simulation results. The presented results also give several insights on the performance of OSA networks with mixed traffic.
We consider in the numerical evaluation a system with two classes of traffic: high priority (HP) and low priority (LP), also denoted as type-1 and type-2 packets, respectively. We assumed exponentially distributed operating periods and considered the cases of exponentially and constantly distributed interruption periods. 
The service time (packet length) is also assumed to have either an exponential or a constant distribution. 

Unless mentioned otherwise, the HP arrival rate is assumed  equal to 0.03 and the average real service times are $E[T_1]=3$ and $E[T_2]=5$ (the unit of time is irrelevant). As an example, if the unit of time is \emph{millisecond} (ms) and the channel data rate is 4 Mbps, those service times represent 1500 bytes and 2500 bytes packets, respectively.

The duration of operating and interruption periods are selected to model two different scenarios. The first scenario is for an almost quasi-static cognitive radio network where $E[Y]\gg E[T]$ and used $E[Y]=75$ and $E[R]=15$.
The second scenario is for a highly dynamic cognitive radio network \cite{khalife09} where $E[Y]<E[T]$ and used $E[Y]=1$ and $E[R]=0.2$. Note that the average server availability is the same for both scenarios, only the dynamics are different.

In the figures, the different distribution cases are denoted by 'ExpExp', 'ExpDet', 'DetExp' and 'DetDet', respectively for the distributions of $T$ and $R$, as summarized in Table \ref{tab:notations2}. The four service disciplines are denoted in the figures as 'Non' (non-preemptive), 'ENo' (exceptional non-preemptive), 'Pr' (preemptive) and 'FP' (preemption in case of failure). The suffix 'Sim' indicates the simulation results, 'The' corresponds to the analytical evaluation of the theoretical results presented in Section~\ref{sec:pr-queueing}, and 'Alt' indicates the analytical evaluation of the theoretical results for the alternative approach presented in Section~\ref{app:alt_approach}.

 \scriptsize
\begin{table}[]
 \renewcommand{\arraystretch}{1.3}
 \caption{{Service time and recovery time distribution cases.}}
 \label{tab:notations2}
 \begin{tabular} {|c|l|l|l|l|}
 \hline
 \hline
\textbf{Service Time ($T$)}&Exponential& Exponential&Constant&Constant\\
 \hline
\textbf{Recovery Time ($R$)}&Exponential& Constant&Exponential&Constant\\
 \hline
\textbf{Notation}&ExpExp& ExpDet&DetExp&DetDet\\
 \hline
\end{tabular}
\normalsize
\end{table}
\normalsize

\subsection{Exponential recovery and service time}
\begin{figure}%
\includegraphics[width=0.5\columnwidth]{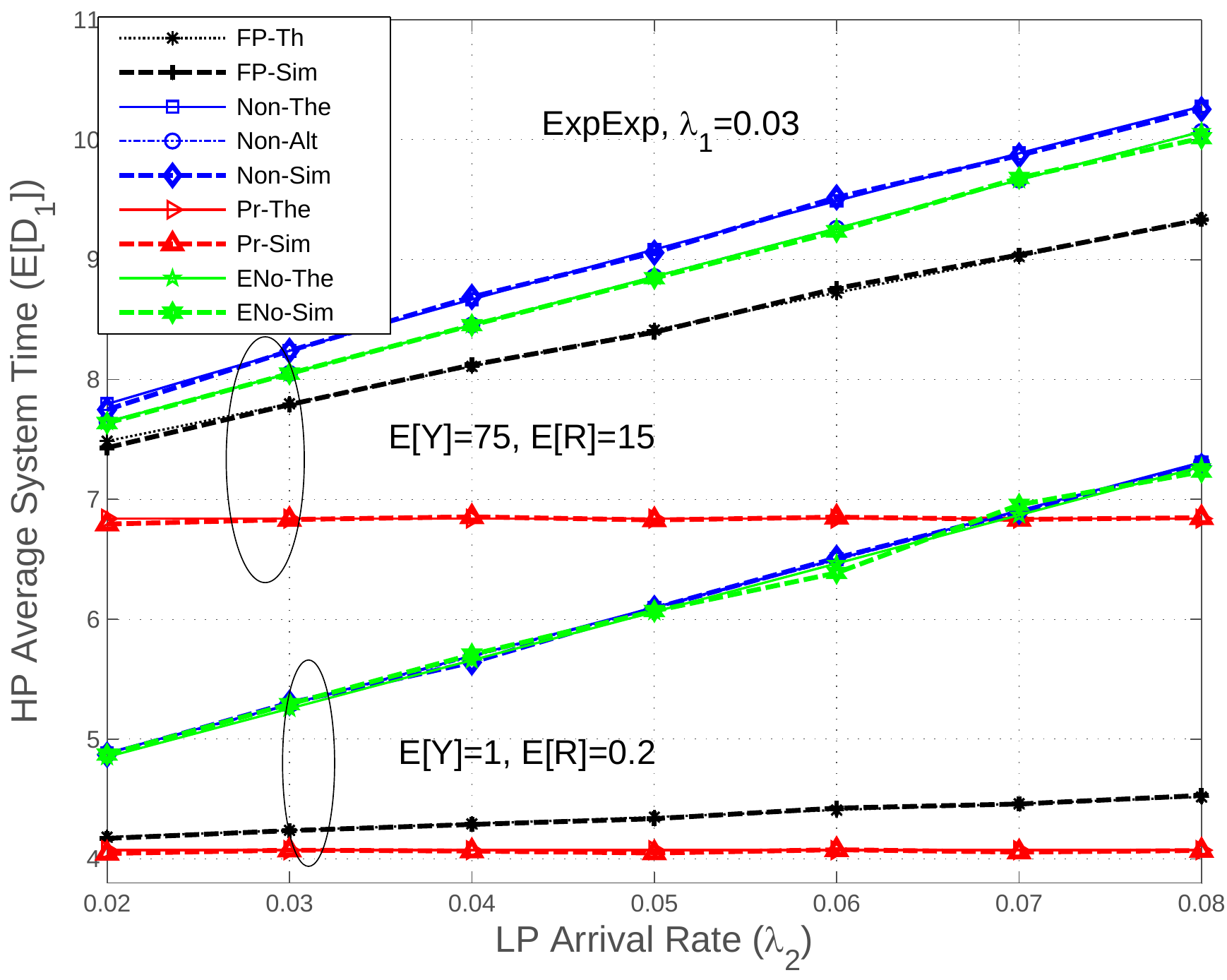}%
\includegraphics[width=0.5\columnwidth]{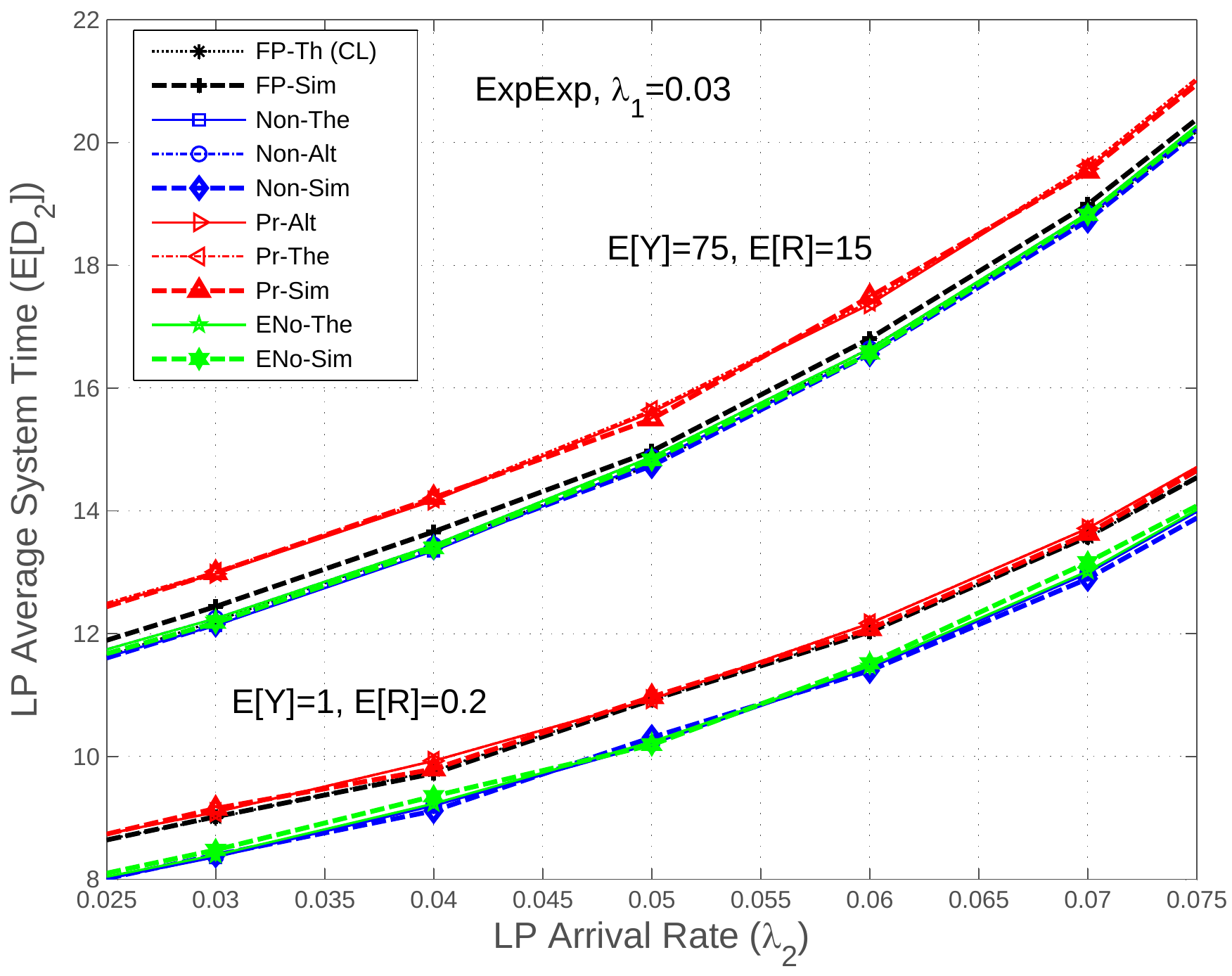}%
\caption{System time of high priority (HP) and low priority (LP) packets vs. LP arrival rate.}%
\label{fig:32212-33212-D1-EE-LargeSmall}%
\end{figure}

The recovery time for a random channel selection recovery algorithm (i.e., the user senses a list of channels one by one until finding an available channel) or a slotted-Aloha competition with other CR users can be accurately modeled with an exponential distribution. 
Fig. \ref{fig:32212-33212-D1-EE-LargeSmall} shows the HP and LP average system time respectively for this case.
As expected, and can also be observed for all the presented results, the HP system time increases from the preemptive, preemptive in case of failure, exceptional non-preemptive and non-preemptive service disciplines, and the LP system time increases in the inverse order of service disciplines. The results also show that the simulation and theoretical analysis results closely match, which validates the priority queueing analysis. 
\begin{figure}%
\includegraphics[scale=0.65]{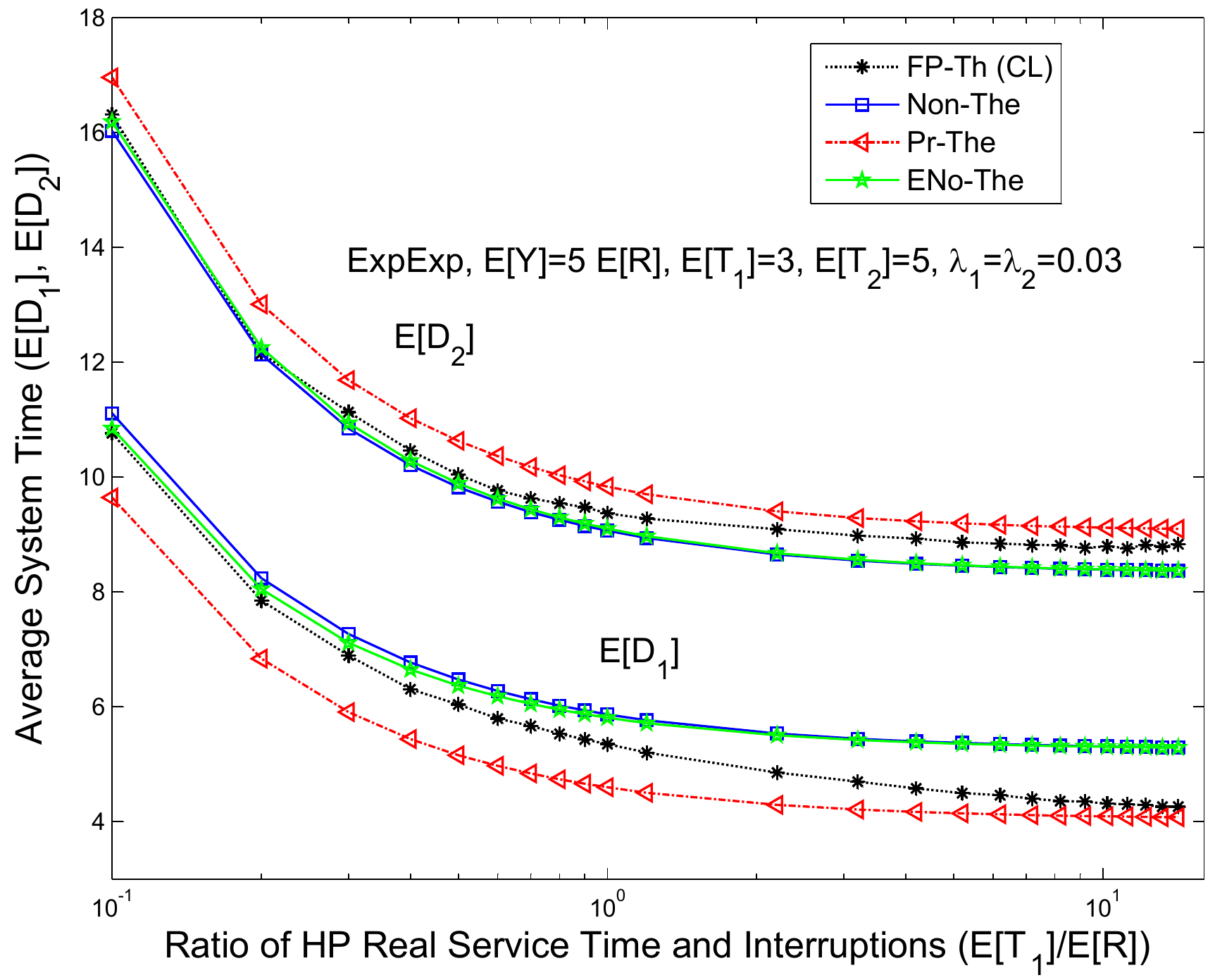}%
\caption{System time of LP and HP packets vs. the variations of $E[R]$ and $E[Y]$ when their ratio is fixed.}%
\label{fig:Fig-fixed-ratio-10092013.pdf}%
\end{figure}

Results are also presented for static and dynamic operating period scenarios.  We can observe that for the same server availability ratio $E[Y]/E[R]$, the system time is worst for both classes of traffic and all service disciplines for the static scenario where $E[R]$ and $E[Y]$ are much larger than the service time. This is due to the fact that the long recovery periods in the static scenario have a severe impact on the OSA queue performance metrics for all traffic classes. To further investigate this important finding, in Figure \ref{fig:Fig-fixed-ratio-10092013.pdf} we present the system time as a function of ${E[T_1]}/{E[R]}$ for a server availability ratio $E[Y]/E[R]$  fixed to five.  This figure clearly shows that the OSA network system time performance deteriorates as the system dynamic decreases (i.e., when ${E[T_1]}/{E[R]}$ decreases) with an inflexion point when the service time is approximately equal to the average interruption length. Furthermore, both HP and LP packets are similarly affected. That is, queueing disciplines can not protect HP traffic against long interruptions. This is expected since interruptions indeed preempt the server. Note that for a traditional OSA throughput analysis based on a saturated-traffic model, no major performance changes will be observed as a function of the OSA network dynamic since the main factor is the server availability ratio $E[Y]/E[R]$. Only the complete queueing analysis presented in this paper can give an insight on the important impact of system dynamics on the OSA performance.

The other interesting point to observe is that for dynamic scenarios, due to frequent short interruptions, the preemption in case of failure service discipline enables the quick preemption of LP packets by HP packets. This service discipline performance is thus close to the preemptive scheme for dynamic scenarios. On the other hand, for static scenarios the preemption in case of failure service discipline performance gets closer to the non-preemptive discipline due to the lack of opportunities for HP packets to preempt LP packet service. 
Meanwhile, the system time for the exceptional non-preemptive discipline is very close to the non-preemptive scheme performance in dynamic scenarios because the probability of an HP arrival in an empty system in the same recovery period as an LP arrival is very low. In large scenarios, their performances start to differ. However, the performance gain remains small. Those results show that the novel priority discipline of preemption in case of failure for OSA networks can significantly improve the system time of HP packets in several deployment scenarios while having a lower implementation complexity than a full preemptive service discipline.

\begin{figure}%
\includegraphics[width=0.5\columnwidth]{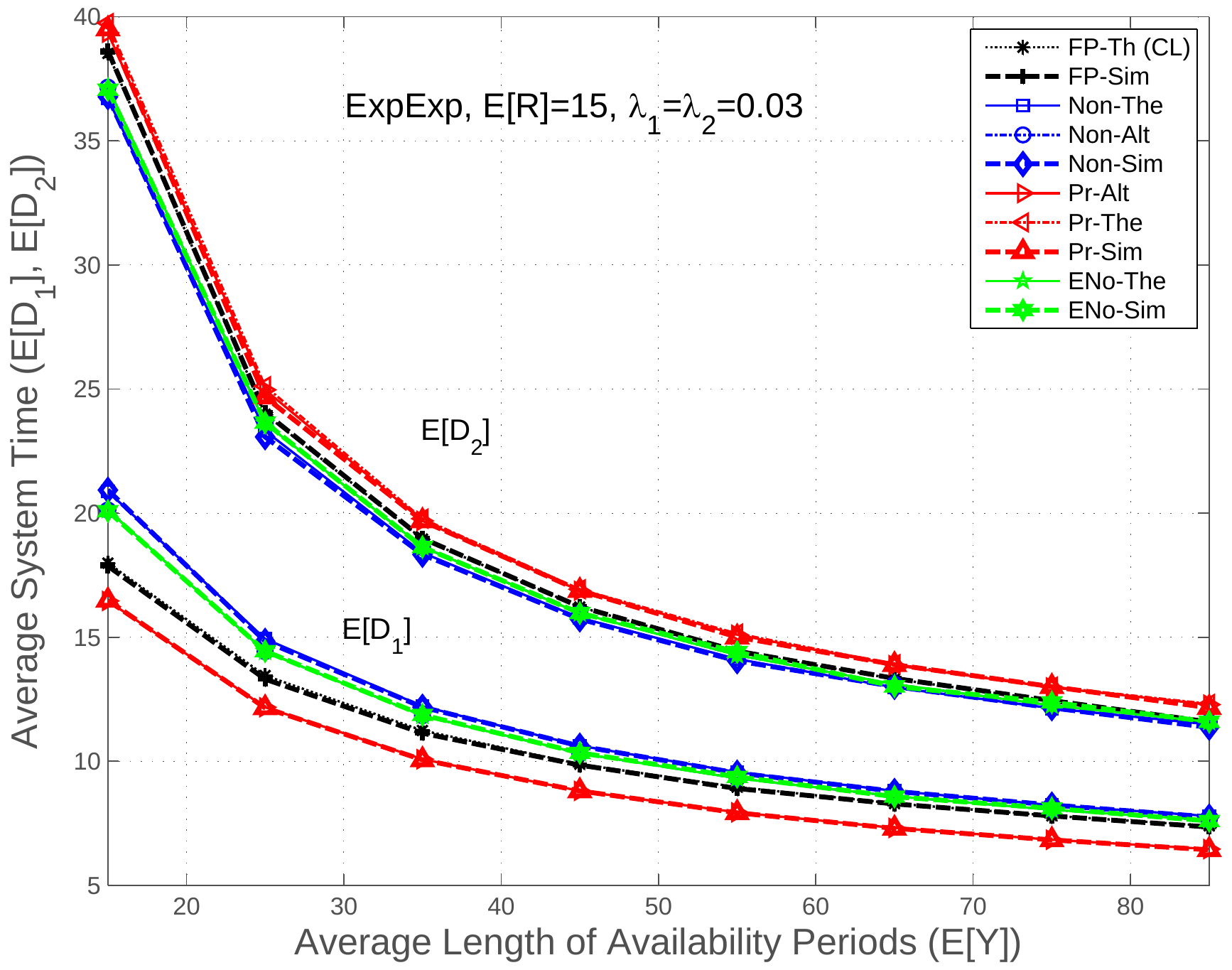}%
\includegraphics[width=0.5\columnwidth]{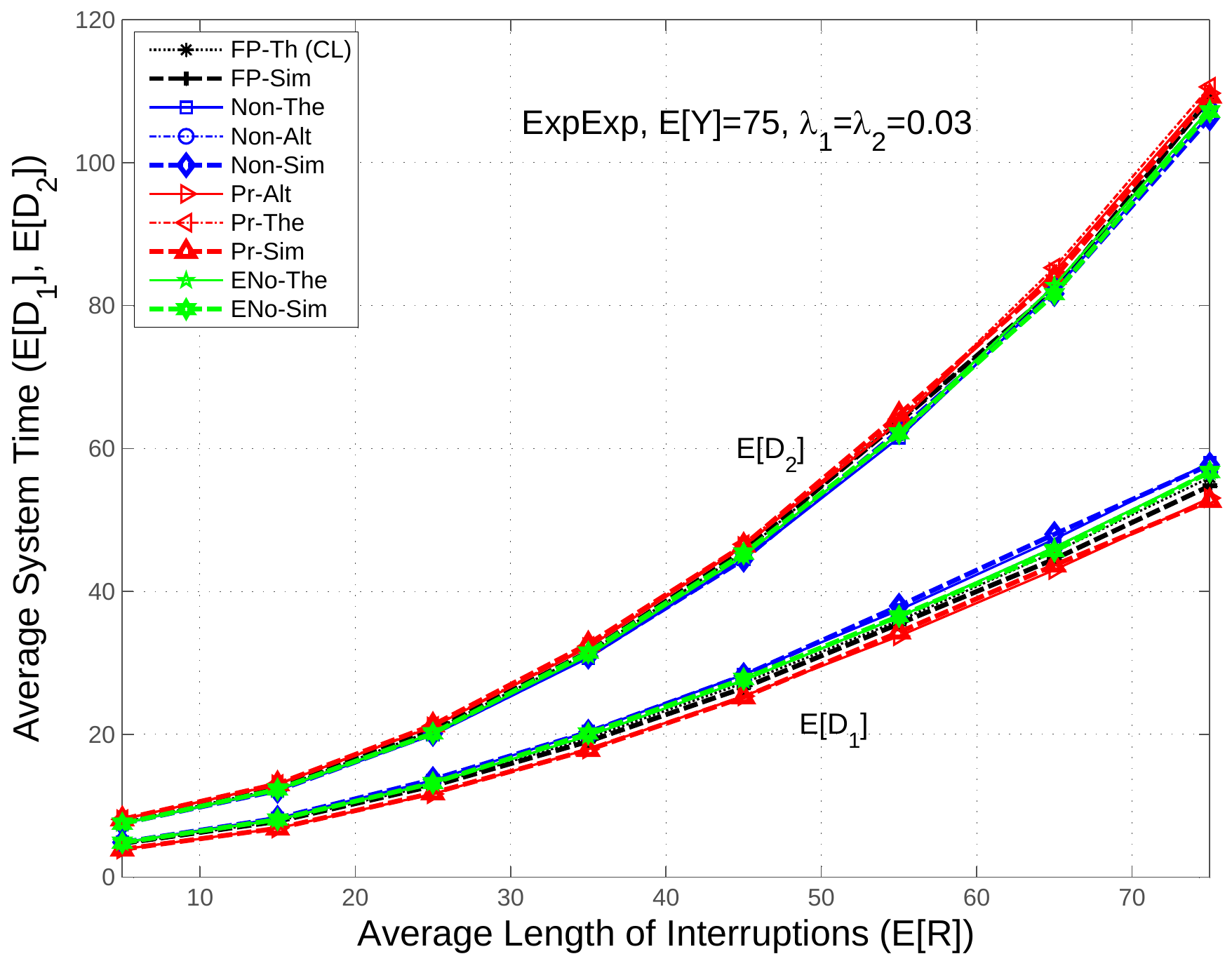}%
\caption{System time of HP and LP packets vs.  operating and interruption period duration.}%
\label{fig:34212-D1D2-EE-Alpha}%
\end{figure}

In Fig. \ref{fig:34212-D1D2-EE-Alpha}, we study the CR traffic system time as a function of the operating and interruption period length. Those results illustrate the validity of the queue analysis for a wide range of operating and interruption periods.  Note that since $E[R]$ or $E[Y]$ is fixed, the server availability increases as a function of $E[Y]$ in the former case and decreases as a function of $E[R]$ in the later case. We can observe that as the server availabitily decreases, either due to shorter availability periods or longer interruption periods, the system time significantly increases, with LP traffic being more affected than HP traffic due to the priority service disciplines. It is interesting to again note that the system time increases much faster when the interruption period increases than when the availability period decreases. For example, starting from the point where $E[Y]=75$ and $E[R]=15$ to the point where $E[Y]=E[R]$, the HP system time approximately increases by a factor of three when the operating period length decreases and by a factor of eight when the interruption period length increases. Those results underline the critical importance of minimizing the interruption period length in OSA networks.

\begin{figure}%
\includegraphics[scale=0.65]{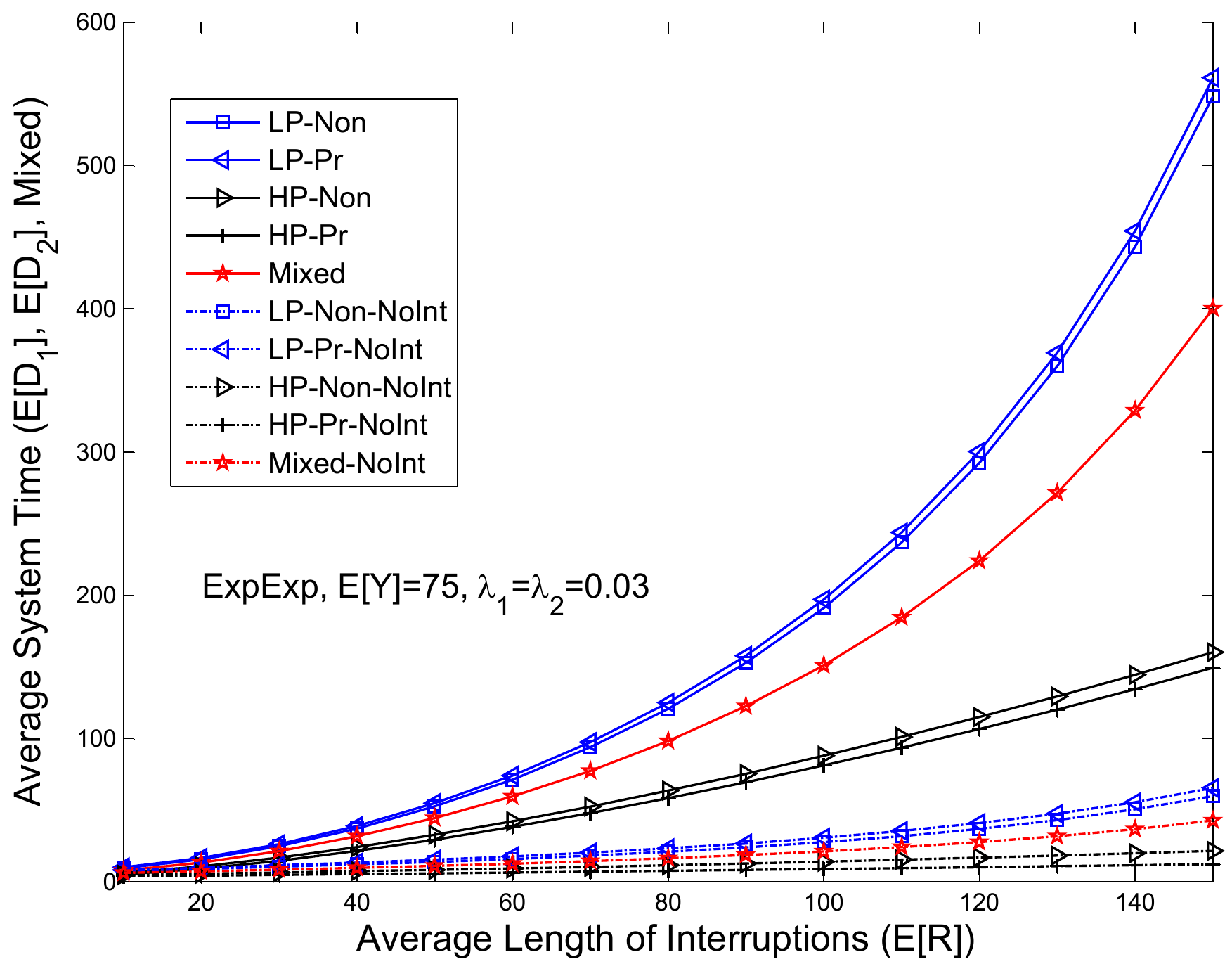}%
\caption{Performance comparison of accurate queueing models with interruption with a modified M/G/1 with longer packets but no interruption.}%
\label{Fig-NoInt-with-Int-3}%
\end{figure}

The results presented in Fig.~\ref{Fig-NoInt-with-Int-3} further motivate the importance of the theoretical analysis provided in this paper to correctly analyze the performance of OSA networks. A straightforward tempting simplification that could be used to analyze the CR queue system time is to use the standard M/G/1 formulas without interruption and increase the packets real service time  $T_i$ by the ratio $\frac{E[Y]+E[R]}{E[Y]}$ to compensate for the average throughput loss due to interruptions. It can be shown for both traffic classes that the queues will saturate at the same traffic load for both the simplified analysis and the correct analysis. However, as can be seen in Fig.~\ref{Fig-NoInt-with-Int-3}, the M/G/1 simplification (referred as 'NoInt') significantly underestimates by almost an order of magnitude the real performance of the queue for both preemptive and non-preemptive priority disciplines (this simplification does not allow the analysis of the two other OSA service disciplines due to the absence of interruptions). This error is due to the fact that the simplified modeling is in fact equivalent to assuming that the interruption periods approach a length of zero. But, as we have discussed previously, interruption periods have a major impact on the OSA queue performance. The accurate modeling of the interruption periods, as we provided in this paper, is thus critical to obtain a valid OSA queue analysis.

Fig.~\ref{Fig-NoInt-with-Int-3} also presents the system time when both traffic classes are mixed without a priority service discipline (i.e., both packet types are queued together and are served in a first-in-first-serve scheme). It is interesting to observe that for a standard system without interruption, the system time increase for LP packets when a priority service discipline is used is almost the same as the system time decrease for HP packets (i.e., the mixed traffic service time is almost exactly in the middle between the LP and HP service times with both priority service disciplines). However, this is not the case for the OSA network where the HP packet system time decreases significantly more than the system time increase for LP packets. This is due to the fact that the interruption periods preempt both classes of traffic. Those results indicate that differentiated service with priority queueing has a bigger impact in  OSA networks than in conventional networks and should thus be actively used when they carry multiple classes of traffic with different QoS requirements.

\subsection{Exponential recovery time and constant service time}

\begin{figure}%
\includegraphics[scale=0.65]{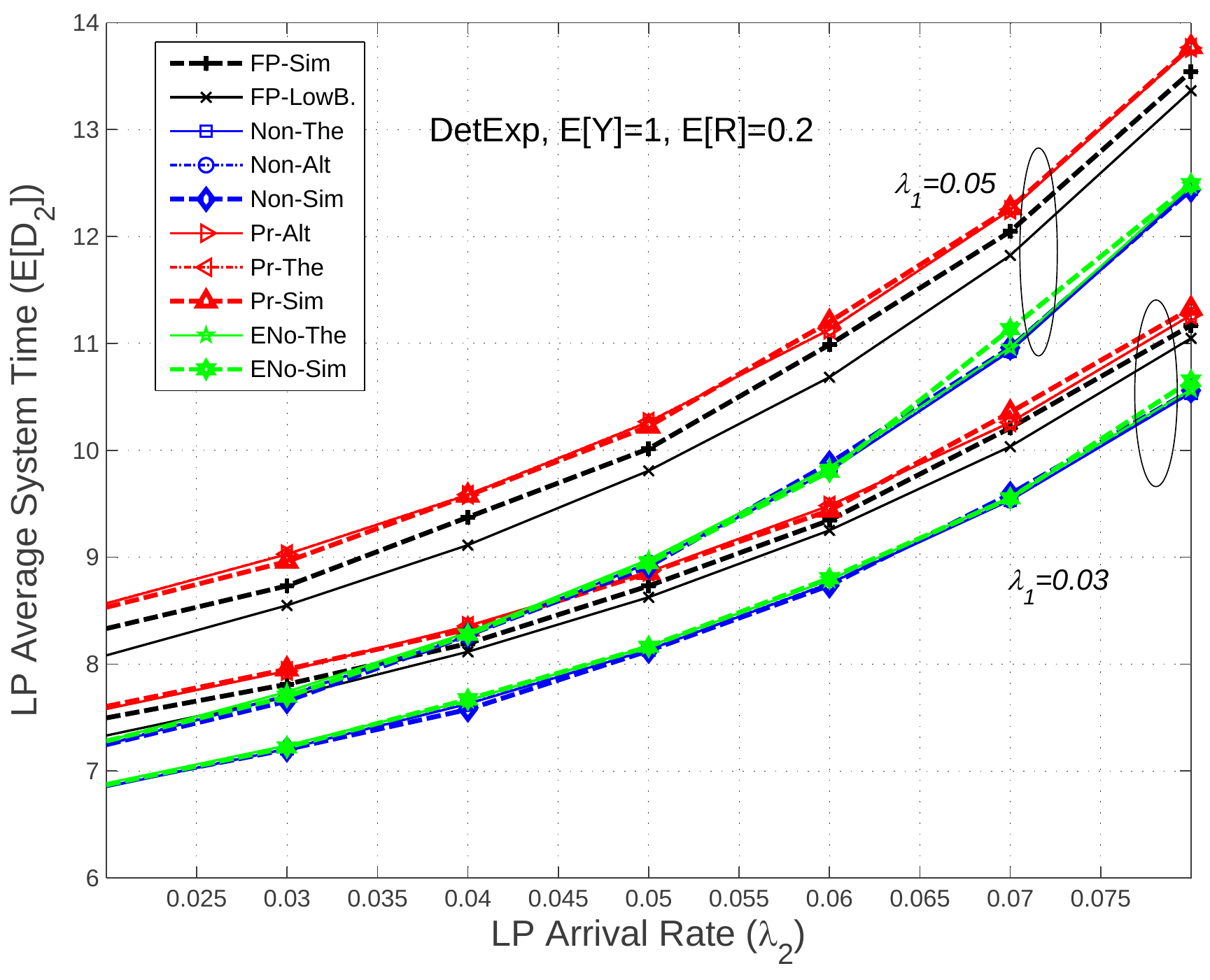}%
\caption{System time of LP packets vs. LP arrival rate for two values of HP arrival rate (small scenario).}%
\label{29212-27212-DESmall}%
\end{figure}

In Fig. \ref{29212-27212-DESmall}, the HP and LP packet lengths are both constant ('DetExp' scenario). Thus, the lower bound for the waiting time of LP packets for the preemptive in case of failure service discipline is presented. The results are generated for two different values of HP arrival rate. To better observe the accuracy of the completion time approximations for the preemptive in case of failure service discipline, simulation results and approximations for the moments of the LP packets completion time, $X_2$ and $X^*_2$ , are compared in the upper part of Table \ref{tab:X2moments-NewPr}. Note that the completion time of LP packets is independent of their arrival rate. The results show the accuracy of the analysis for non-exponential service time and the slight deterioration due to the approximate analysis for the preemptive in case of failure service discipline. On the other hand, we can observe the accuracy of the theoretical analysis with deterministic packet service time for the three other service disciplines.

\subsection{Constant recovery time}
A constant recovery time occurs, for instance, in scenarios where the information concerning the channels' occupancy is provided in advance; therefore, no random sensing is required and the recovery time only represents a constant time for negotiation and radio alignment. In order to compare the results with the previous scenario, we assume the same average values. 

\begin{figure}%
\includegraphics[width=0.5\columnwidth]{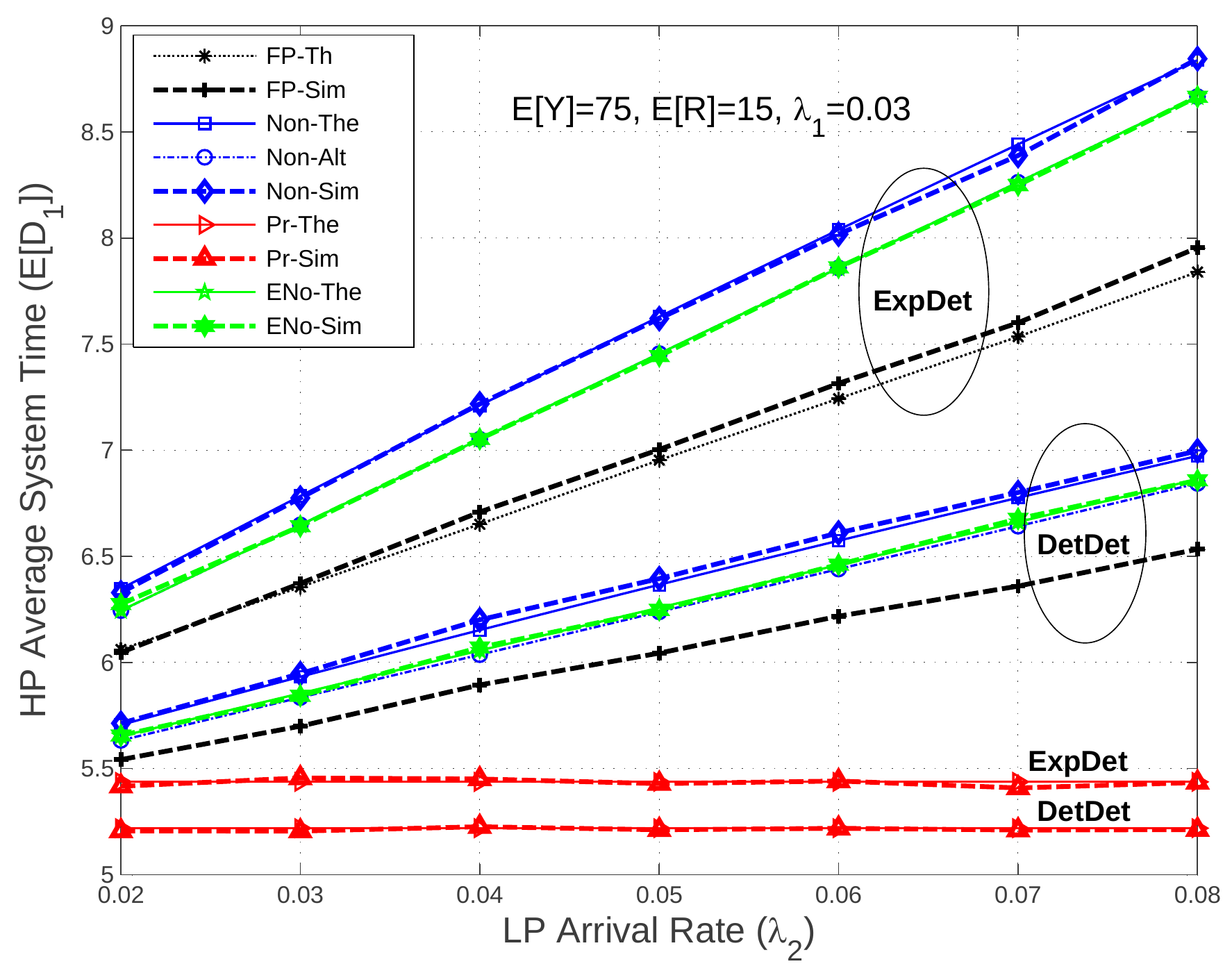}%
\includegraphics[width=0.5\columnwidth]{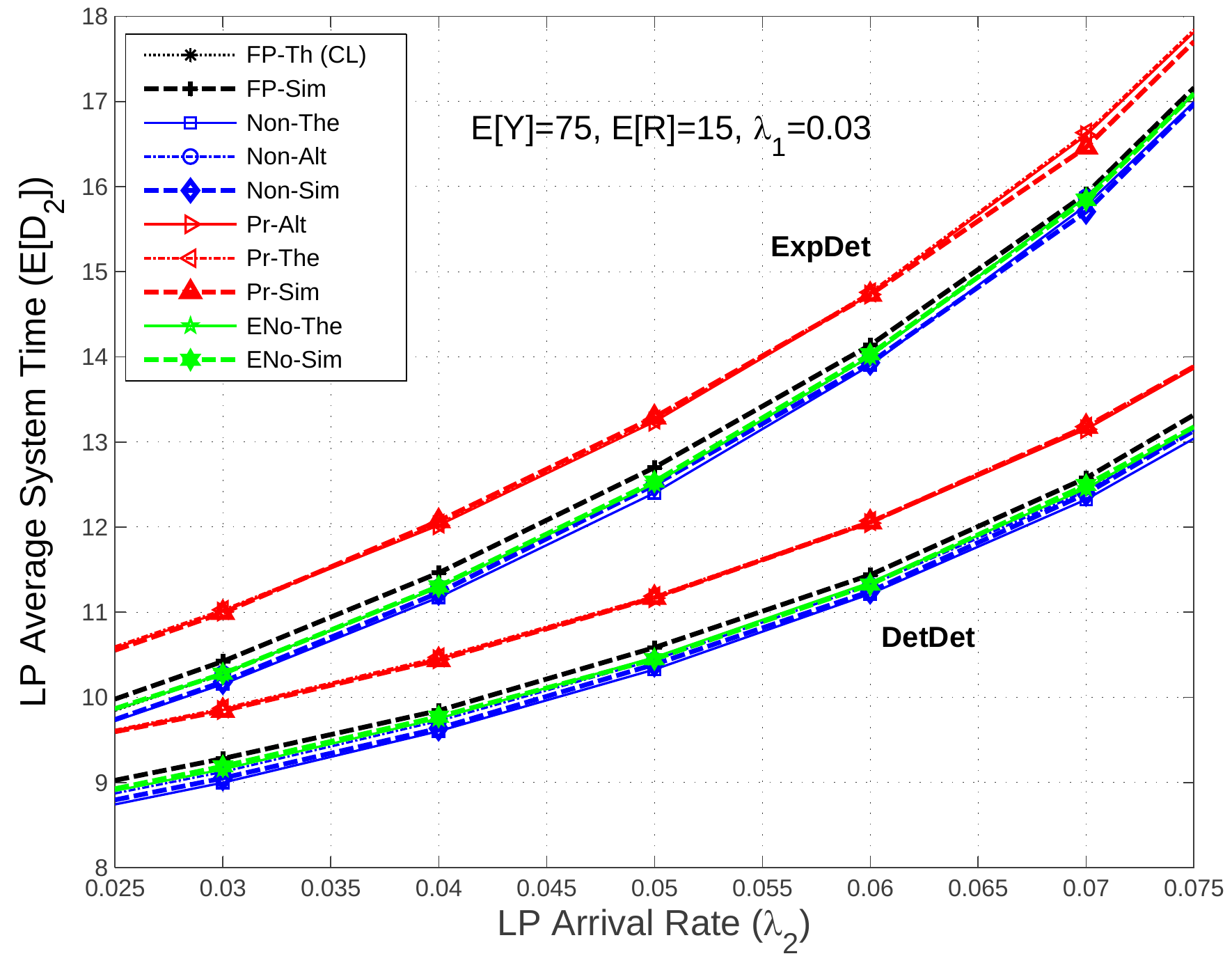}%
\caption{System time of high (HP) and low priority (LP) packets vs. LP arrival rate (DetDet and ExpDet, large scenario).}%
\label{30212-D2-ED-Large}%
\end{figure}

Fig. \ref{30212-D2-ED-Large} illustrates the system time of HP and LP packets versus their arrival rate for the cases of exponential and constant service times. 
It should be noted that for the selected values in the simulation, the upper and lower bounds proposed in Section \ref{subsec:prf-Waiting-time} (e.g., Eq. (\ref{eq:eq:waiting-with-vacation2-HP})) are loose compared to the natural bounds of preemptive and non-preemptive disciplines, so they are not illustrated in the figures to enhance their clarity.
As expected, it can be seen in both figures that the performance, when the distribution of real service time (packet length) is exponential, is worse compared to the case where the packet length is constant (with the same average). The results also validate the theoretical analysis presented in Section~\ref{sec:pr-queueing}. However, it can be observed that the system time for the non-preemptive service discipline obtained with the alternative approach presented in Section~\ref{app:alt_approach} is not accurate. This shows, as discussed in the introduction, the limitations of alternative approaches which were previously proposed in the literature when the recovery period is not exponentially distributed. Furthermore, this alternative approach can not be used to analyze more sophisticated service disciplines such as the exceptional non-preemptive and the preemptive in case of failure disciplines which provide interesting performance gains for OSA networks.
Simulation and analytical results for the completion time of LP packets are compared in the lower part of Table \ref{tab:X2moments-NewPr} for two different values of HP arrival rate. 
\begin{table}[]
 \renewcommand{\arraystretch}{1.3}
 \scriptsize
 \caption{{Moments of the LP completion time with preemption in case of failure (FP) service discipline for DetExp and DetDet scenarios (S: Simulation, A: Approximation). }}
 \label{tab:X2moments-NewPr}
 \begin{tabular} {|c|c|l|l|l|l|}
 \hline
 {\textbf{Scen.}}&{\textbf {$\lambda_1$}}&{\textbf {$E[X_2]$-S}} &{\textbf {$E[X_2]$-A}}&{\textbf {$E[X^2_2]$-S}}&{\textbf {$E[X^2_2]$-A}} \\
 \hline
Small,DE&0.03&6.62&6.72&47.01&49.51\\
 \hline
Small,DE&0.05&7.10&7.32&57.05&62.33\\
 \hline
Large,DE&0.03&6.13&6.12&77.53&74.96\\
 \hline
Large,DE&0.05&6.26&6.25&88.37&85.62\\
\hline
\hline
Small,DD&0.03&6.61&6.72&46.76&49.23\\
 \hline
Small,DD&0.05&7.10&7.31&56.69&61.98\\
 \hline
Large,DD&0.03&6.14&6.12&58.40&56.45\\
 \hline
Large,DD&0.05&6.26&6.25&65.00&63.64\\
\hline
\end{tabular}
\normalsize
\end{table}
\normalsize

\section{Conclusion and Future Work}
\label{eq:conclusion}
Priority queueing is a classical approach to implement traffic differentiation in communication links. To analyze priority queueing schemes for opportunistic spectrum access networks, we derived in this paper a general queueing model
with interruptions for the preemptive and non-preemptive classical priority
disciplines. Two new cognitive radio disciplines were also introduced in this paper: exceptional non-preemptive and  preemptive in case of failure. The theoretical analysis was validated with simulation results and we investigated the behavior of those disciplines for different sets of parameters and distributions for the packet service time and interruption periods. We also showed how the analysis can be used by an OSA controller to make critical decisions such as selecting the channel switching policy or the priority queueing discipline based on the estimated channel parameters. It was also observed that even though the ratio of operating and interruption periods plays an important role, a significant performance decrease is observed in a semi-static network with long operating and interruption periods compared to a fast-varying network with short periods and the same ratio. A simplified M/G/1 model with no interruption and compensated increased packet length can not thus capture the queue metrics performance of OSA networks. We also presented results demonstrating the importance of priority queueing to provide differentiated service in the presence of frequent interruptions. 
As discussed in the introduction, an important area of future work is to use the results presented in this paper to further study and optimize MAC protocols and channel assignment policies in OSA networks based on not only a saturated mode throughput analysis but also on queue metrics.
Another interesting area of research is to extend this work to the cases of queueing with service repeat after an interruption and for non-homogeneous channels with variable service rate for different operating periods.

\bibliography{new-refs}
\bibliographystyle{IEEEtran}

\end{document}